\documentclass[11pt,a4paper]{article}

\usepackage{jcappub}

\usepackage{graphicx}
\usepackage{enumitem}
\usepackage{amsmath}
\usepackage{booktabs}
\usepackage{array}
\usepackage{color}
\usepackage{amssymb}
\usepackage{mathrsfs}
\usepackage[T1]{fontenc}
\usepackage[latin1]{inputenc}
\usepackage[table]{xcolor}  
\usepackage{verbatim}
\usepackage{caption}
\usepackage{subcaption}
\usepackage{rotating}
\usepackage{dcolumn}
\usepackage{bm}

\renewcommand{\d}{\mathrm{d}}

\newcommand{\ipleft}{\langle\kern-0.2em\langle}
\newcommand{\ipright}{\rangle\kern-0.2em\rangle}

\newcommand{\bx}{{\mathbf{x}} }
\newcommand{\bn}{{\mathbf{n}} }
\newcommand{\bk}{{\mathbf{k}} }
\newcommand{\GG}{{\mathcal{G}} }

\newcommand{\Ll}{{\mathcal{L}} }
\newcommand{\tba}{{\overline{t}} }
\newcommand{\vba}{{\overline{v}} }

\newcommand{\Ktwo}{{\widehat{{k}}_3} }

\newcommand{\para}[1]{\par\vspace{2mm}\noindent\textit{{#1}---}}

\newcolumntype{s}{>{$\displaystyle}l<{$}}
\newcolumntype{t}{>{$\displaystyle}c<{$}}
\newcolumntype{u}{>{$\displaystyle}r<{$}}
\newcolumntype{v}{>{$\displaystyle}m{4cm}<{$}}

\newcolumntype{d}{D{!}{\;\pm\;}{-1}}

\def\al{\alpha} 
\def\be{\beta} 
\def\ga{\gamma}
\def\de{\delta}
\def\ep{\epsilon}

\def\et{\eta}

\def\ka{\kappa}

\def\Th{\Theta}

\def\Om{\Omega}

\newcommand{\ben}{\begin{equation}}
\newcommand{\een}{\end{equation}}
\newcommand{\bea}{\begin{eqnarray}}
\newcommand{\eea}{\end{eqnarray}}
\newcommand{\ba}{\begin{array}}
\newcommand{\ea}{\end{array}}
\newcommand{\bit}{\begin{itemize}}
\newcommand{\eit}{\end{itemize}}
\newcommand{\vev}[1]{\left\langle#1\right\rangle}

\newcommand{\erf}{{\text{erf}}}
\newcommand{\cA}{{\cal A}}

\newcommand{\etaEq}{\eta_\text{eq}}
\newcommand{\zEq}{z_\text{eq}}
\newcommand{\PpostEq}{P}
\newcommand{\BpostEq}{B}

\newcommand{\Xs}{X}
\newcommand{\bXs}{\mathbf{X}}
\newcommand{\cL}{{\cal L}}
\newcommand{\nwake}{n_\text{w}}

\newcommand{\etaInit}{\eta_\text{i}}

\newcommand{\fd}{f_{10}}
\newcommand{\uetc}{UETC}
\newcommand{\cmbact}{CMBACT}

\def\setsize{\csname @setfontsize\endcsname \setsize}

\title{The  
bispectrum of matter perturbations from cosmic strings}
\author[1]{Donough Regan}
\author[1,2]{and Mark Hindmarsh}
\affiliation[1]{Astronomy Centre, University of Sussex,
Falmer, Brighton, BN1 9QH, UK}
\affiliation[2]{Department of Physics and Helsinki Institute of Physics, P.O.\ Box 64, 00014 Helsinki University, Finland} 

\emailAdd{d.regan@sussex.ac.uk}
\emailAdd{m.b.hindmarsh@sussex.ac.uk}
\abstract{We present the first calculation of the bispectrum of the matter perturbations induced by cosmic strings. 
The calculation is performed in two different ways: the first uses the unequal time correlators (UETCs) of the string network - computed using a Gaussian model previously employed for cosmic string power spectra. The second approach uses the wake model, where string density perturbations are concentrated in sheet-like structures whose surface density grows with time. 
The qualitative and quantitative agreement of the two gives confidence to the results.
An essential ingredient in the \uetc\ approach is the inclusion of compensation factors in the integration with the Green's function of the matter and radiation fluids, and we show that these compensation factors must be included in the wake model also. We also present a comparison of the UETCs computed in the Gaussian model, and those computed in the unconnected segment model (USM) used by the standard cosmic string perturbation package CMBACT. 
We compare numerical estimates for the 
bispectrum of cosmic strings to those produced by perturbations from an inflationary era, and discover that, despite the intrinsically non-Gaussian nature of string-induced perturbations, the matter bispectrum is unlikely to produce competitive constraints on a population of cosmic strings.
}

\begin{document}	
\maketitle

\section{Introduction}
Cosmic strings are linear topological defects formed after a symmetry-breaking phase transition in the early Universe \cite{Kibble:1976sj,VilShe94,Hindmarsh:1994re,Copeland:2011dx,Hindmarsh:2011qj}. 
They are characterised by their dimensionless mass per unit length, $G\mu$, where $G$ is Newton's constant, and the mass per unit length $\mu$ is proportional to the square of the 
symmetry breaking scale.
Simulations and analytical modelling in an expanding universe show that a network of cosmic strings relaxes towards a so-called scaling regime, 
in which the average string separation is of order the Hubble length. 
However the mechanism by which the cosmic string network loses energy - typically via the transfer of energy from large scales to smaller scales - is still not well understood. In the Nambu-Goto approximation, where strings are modelled with zero width, energy is transferred from ``infinite'' (longer than the horizon length) strings to smaller loops by reconnection, and then presumably into gravitational radiation \cite{Albrecht:1989mk,Bennett:1989yp,Allen:1990tv,Ringeval:2005kr,Martins:2005es,BlancoPillado:2011dq,Blanco-Pillado:2013qja}.
In simulations of an underlying Abelian Higgs field theory \cite{Vincent:1997cx,Moore:2001px,Bevis:2006mj, Hindmarsh:2008dw,Bevis:2010gj}, the energy is lost through the production of classical radiation of the fields. 
Which of these models is more accurate at late times is unknown, and may be model-dependent.


There are strong and largely model-independent constraints from the Cosmic Microwave Background (CMB) power spectrum. The Planck collaboration gives 95\% confidence upper limits of $G\mu < 1.3\times 10^{-7}$ in the Unconnected Segment Model (USM) of the Nambu-Goto (NG) scenario, and  $G\mu < 3.2\times 10^{-7} $ in the Abelian Higgs model \cite{PlanckCosmicStrings}.  Slightly stronger limits can be imposed if BICEP2 data is included \cite{Lizarraga:2014xza}. The constraints are often also quoted in terms of $\fd$, the fractional contribution of strings to the temperature power spectrum at a multipole $\ell=10$.  In the Abelian Higgs model, the Pplnack upper bound is $\fd < 0.028$ with 95\% confidence.

Perturbations from strings are expected to be much more non-Gaussian than those generated by inflation, particularly on small scales, as they are spatially localised. Limits on the bi- and tri-spectra of the CMB perturbations can be used to constrain $G\mu$ \cite{HRS09x1,HRS09x2,ReganShellard09}, and the Planck CMB bispectrum data results in a conservative constraint of $G\mu \lesssim 8.8 \times 10^{-7}$. The possibility of using the CMB trispectrum was investigated in \cite{FRS10}. However, the signal is very difficult to measure and the constraints produced are weak.

In this paper we produce the first calculation of the cosmic string matter bispectrum, which appears a promising route for constraints or detection as, unlike the CMB bispectrum, it is not suppressed by symmetry considerations. Our method introduces the unequal time 3-point correlator of the source function (\uetc3). We integrate it with the Green's functions for a simplified universe consisting of dark matter and radiation only.  
Our model for the \uetc3 is based on a Gaussian approximation to the string ensemble, successfully used for calculations of the cosmic string CMB power spectrum \cite{Hindmarsh:1993pu} and higher order correlators \cite{ReganShellard09}. 

We compare our results with the wake model for cosmic string matter perturbations \cite{Silk:1984xk,Vachaspati:1986zz,Stebbins:1987cy,VilShe94}, in which the perturbations are entirely in the form of sheet-like structures with a surface density growing according to the Zel'dovich approximation. 
We show that the bispectrum can be understood in terms of a random superposition of wakes, provided proper attention is paid to the issue of compensation \cite{Veeraraghavan:1990yd}.

As the method is new, we validate it by calculating the cosmic string matter power spectrum \cite{MelottScherrer87,Albrecht:1991br,Avelino:1995zm,Avelino:1997hy,Avelino:1998qd,CHM98,Avelino:1999jb,Wu2002} 
and comparing it to that produced by \cmbact\ \cite{CMBACT,CMBACTx2}, a standard package for computing string perturbation power spectra, via an implementation of the USM \cite{Albrecht:1997mz}. 
We also compare our 2-point unequal time correlator (\uetc) computed in the underlying Gaussian string model to a recent analytic computation of the \uetc\ captured by \cmbact\ \cite{12092461}.

We refer the reader to related work in \cite{Jaffe:1993tt,Figueroa:2010zx} on the matter bispectrum in the context of non-topological defects due to self-ordering scalar fields (SOSFs), modelled as a non-linear sigma model of $N$ scalar fields with $\mathcal{O}(N)$ symmetry. In these studies, it was shown that in the large $N$ limit, the bispectrum may be calculated. The form of the bispectrum on large scales, as well as general considerations in \cite{Pen:1997ae}, suggest that the matter bispectrum is unlikely to provide competitive constraints to the CMB bispectrum of SOSFs, with the contribution due to non-linear evolution of inflationary perturbations (expected to be Gaussian, adiabatic, and approximately scale-free) dominating on small scales. However, the power spectrum of perturbations from strings decreases much less rapidly at small scales than either SOSFs or inflationary perturbations, and the perturbations from strings are intrinsically non-linear and non-Gaussian from the outset. 
The formalism developed for SOSFs is not directly applicable, and different techniques are applied in this paper. Nevertheless, as we shall see, the general features of the large scale matter bispectrum are similar, and the conclusion of the bispectrum's potential detectability -- even at small scales -- remains the same.

In Section~\ref{sec:densityperturbation} we describe our \uetc-based method to estimate the matter perturbation due to cosmic strings, and give a review of the wake model. 
In Section~\ref{sec:compensation}, we discuss compensation, an essential component of both models.
We apply each of the models to the computation of the power spectrum in Section~\ref{sec:powerspecSEC}. We compare the results of each of the approaches in certain asymptotic limits, verifying that they obey the same qualitative behaviour. We provide a comparison to the output of \cmbact. 

Finding excellent agreement between all three calculations, we develop the \uetc3 approach to the matter bispectrum in Section~\ref{sec:bispecSEC}, and compare it to the wake model. In both cases we provide analytic estimates for the qualitative behaviour in the equilateral, squeezed, and folded configurations in $k$-space. 
We also compare the matter bispectrum of cosmic strings to that induced by gravitational effects at second order, and discuss whether such a signal may be detectable. Finally in Section~\ref{sec:conclusions} we present our concluding remarks.



\section{Density perturbations from cosmic strings}\label{sec:densityperturbation}
In this section two approaches are described for the evaluation of the matter density perturbation due to a network of cosmic strings. First we describe the linear perturbation theory calculation. The statistics of perturbations due to cosmic strings are generally evaluated in this framework, since they must describe a subdominant contribution to the power spectrum. However, it is not a priori clear whether this remains true at the level of the three-point correlator (bispectrum). Therefore, we also present a non-linear estimate for the perturbation, which we term the wake model. It uses the Zel'dovich approximation to describe the wake-like accumulation of matter behind a moving cosmic string. 

\subsection{Linear perturbation theory}\label{subset:pertTheory}
We consider a universe consisting of radiation and cold dark matter (CDM) only, which is an adequate approximation at our level of modelling. 
The equations of motion for the radiation and CDM density perturbations $\delta_{r,c}$ in the synchronous gauge can be written as \cite{VeerStebbins,9712008,Wu2002}
\begin{align}\label{eq:linevol}
\ddot{\delta_c}+\frac{\dot{a}}{a}\dot{\delta_c}-\frac{3}{2} \left(\frac{\dot{a}}{a} \right)^2 \left(\frac{a\delta_c +2 a_{{\rm{eq}}}  \delta_r      }{a+ a_{{\rm{eq}}} +\frac{\Omega_{\Lambda}}{\Omega_{c}  }  \frac{a^4}{a_0^3}      } \right)&=4\pi G \Theta_+,\\
\ddot{\delta_r}-\frac{1}{3}\nabla^2 \delta_r -\frac{4}{3}\ddot{\delta_c}=0,
\end{align}
where the source term due to the string energy-momentum tensor is given by $\Theta_+=\Theta_{00}+ \Theta_{ii}$, $a$ is the scale factor (with the subscript $\rm{eq}$ denoting matter-radiation equality, and $0$ denoting today), and a dot represents differentiation with respect to the conformal time, $\eta$. Note that in order to write a choice (\ref{eq:linevol}) a choice of coordinates is made such that the CDM velocity perturbation is zero.


We can solve the equations with a Green's function technique, with initial conditions set at some arbitrary time $\eta_i$ much smaller than any of interest (and certainly much smaller than $\etaEq$, the time of equal matter and radiation density).  We denote the homogeneous part of the solution by $\delta_N^I$, where $N=c,r$, and the particular solution as $\delta_N^S$.  The total density perturbations $\delta_N=\delta_N^I+\delta_N^S$ and the string source $\Theta_{00}$ together satisfy an energy density conservation equation. This so-called compensation \cite{Veeraraghavan:1990yd} is discussed in the next section.

The Green's functions $\mathcal{G}_N$ satisfy $\mathcal{G}_c=0=\mathcal{G}_r,\,\, \dot{\mathcal{G}_c}=1=3\dot{\mathcal{G}_r}/4$ at $\eta=\eta'$, 
and $\mathcal{G}_N=0$ for $\eta<\eta'$), and in Fourier space the particular solution is 
\begin{align}\label{deltaNsol}
\delta_N^S({\mathbf{k}},\eta)=4\pi G\int_{\eta_i}^{\eta} d\eta'\mathcal{G}_N(k;\eta,\eta') \Theta_+({\mathbf{k}},\eta').
\end{align}
As shall be described in Section~\ref{sec:compensation} it is in practice necessary to include a multiplicative (compensation) factor in order to account for the initial perturbation, which ensures that energy momentum conservation is accounted for on super-horizon scales \cite{Albrecht:1991br}. The scheme can be justified empirically by comparison of the resulting power spectrum with that produced by numerical codes which directly integrate the equations of motion. Despite this slight drawback, we have utilised the Green's function approach in order to readily produce analytic estimates.

The energy-momentum tensor due to a cosmic string with spacetime trajectory $X^{\mu}_s=(\eta,{\mathbf{X}}(\sigma,\eta))$ where $(\sigma,\eta)$ denotes the string worldsheet coordinates, is given by
\begin{align}
\Theta_{\mu \nu}(\bx,\eta)=\mu \int d\sigma \left( \epsilon_s \dot{X}^{\mu} \dot{X}^{\nu}-\epsilon^{-1} {X'}^{\mu} {X'}^{\nu}\right) \delta^{(3)} (\mathbf{x-X})\,, 
\end{align}
where prime denotes derivatives with respect to $\sigma$ and $\epsilon=\sqrt{ {\mathbf{X'}^2}/{( 1-{\dot{\mathbf{X}}^2}) }  } $ and we impose $\dot{\mathbf{X}}\cdot\mathbf{X'}=0$. This implies $\Theta_+(\bx,\eta)=2\mu \int d\sigma \epsilon {\dot{\mathbf{X}}^2}\delta^{(3)} (\mathbf{x-X})$, or in Fourier space
$
\Theta_+(\bk,\eta)=2\mu \int d\sigma \epsilon {\dot{\mathbf{X}}^2} e^{i\mathbf{k}\cdot {\mathbf{X}}(\sigma)}\,.
$
We can change the variable $d\sigma\rightarrow ds= \epsilon d\sigma\implies s\approx|{\bf X}| /\sqrt{1-\dot{{\bf X}}^2}$ such that
\begin{align}\label{eq:thetaplus}
\Theta_+(\bk,\eta)=2\mu \int ds {\dot{\mathbf{X}}^2} e^{i\mathbf{k}\cdot {\mathbf{X}}(s)}\,.
\end{align}
It should be noted that the creation of a cosmic string network results in compensating white noise perturbations in the remaining energy density $\de_N$ on super-horizon scales, as a result of energy and momentum conservation. In Section~\ref{sec:compensation} we will discuss this issue in more detail.


\subsection{Nonlinear approach -- wake model}\label{subsec:wakemodel}
In this section we present an alternative - non-linear - method to obtain an analytic formula for the matter power spectrum due to cosmic strings, based on the wake model \cite{Silk:1984xk,Vachaspati:1986zz,Stebbins:1987cy,VilShe94}.

Our implementation of the wake model assumes the following.
\begin{itemize}
\item There are a constant number of long cosmic strings per Hubble volume.
\item The comoving correlation length, $\xi$, of the cosmic strings formed at conformal time $\eta_i$ is given at all times subsequently by $\xi(\eta,\eta_i)=\alpha \eta_i$, where $\alpha$ is a constant.
\item The physical surface density at time $\eta$ of a wake laid down by a (long straight) cosmic string formed at time $\eta_i$ in the matter era is 
\ben
\sigma_w=\left(\frac{4u_i \eta_i}{5}\right) \left(\frac{\eta}{\eta_i}\right)^2 \rho_{c},
\een
where $\rho_{c}$ is the background matter density and $u_i = 4\pi G \mu v_s$.

\item The wake due to a string formed at $\eta_i$ is modelled in comoving space as a two dimensional disk of radius given by the correlation length of the cosmic string, $\xi(\eta,\eta_i)$.

\item Wakes give the dominant contribution to the matter power spectrum. 

\end{itemize} 

The first two assumptions are due to the well established scaling behaviour of long cosmic strings:
numerical simulations show that the number of long strings per Hubble volume is $\mathcal{O}(10)$, with $\alpha\approx 0.15-0.3$. Nambu-Goto simulations give lower values of $\al$ than field theory simulations.

The third assumption is due to the process through which the wakes are formed. Specifically, long straight cosmic strings moving with velocity $v_s$ 
induce a velocity perturbation $\pm 4\pi G \mu v_s$, pulling matter into a wake behind them. 
The subsequent matter perturbation may be found by solving the Zeldovich equation for the comoving displacement from the central plane (see e.g. \cite{VilShe94}).

Our fourth assumption assumes that the length and the width of the string wake are comparable and much larger than the thickness of the wake. On dimensional grounds we expect that the initial physical length, width and thickness are given by $l_i\sim t_i, w_i\sim v_s t_i$ and $d_i\sim u_i t_i$, respectively (where $t_i$ is the physical time corresponding to $\eta_i$). The corresponding length, width and thickness today are given by $l_0\sim t_i z_i, w_0\sim v_s t_i z_i, d_0 \sim u_i t_i z_i^2$. Therefore, our assumption is valid for $v_s$ relativistic and $u_i z_i \ll 1$, which is true for the range of redshifts we are interested in.  

The final assumption, that wakes due to long straight strings give the dominant contribution to the matter power spectrum, is essentially equivalent to the grounding assumption of the USM, widely used to model the CMB. 
It should be noted that the presence of small-scale structure on the strings is expected to significantly change the baryon abundance in the central regions of the wakes \cite{9608020}.

In order to perform the calculation of the total cosmic string matter power spectrum and bispectrum, we must first calculate the density profile of a single cosmic string wake. Consider a cosmic string wake at time $\eta$, created due to a string formed at $\eta_i$, modelled as a two dimensional disk of radius $\xi(\eta,\eta_i)$ with surface density $\sigma_w(\eta,\eta_i)$. The 2D Fourier transform of the disk is given by
\begin{align}
\tilde{F}(k_{\perp};\xi) = 2\pi \frac{J_1(k_{\perp}\xi(\eta,\eta_i))}{k_{\perp}}\,,
\end{align}
where $k_{\perp}=\sqrt{k_x^2 +k_y^2}$ and $J_1$ represents the Bessel function of the first kind (of order one).
We suppose that the string wake is oriented in direction $\hat{\bf{n}}$, which we assume for simplicity is aligned with the $z$-axis. The Fourier transform (FT) of the density profile reveals 
\begin{align}\label{eq:delta1}
\delta_1(\bk,\eta,\eta_i)=\int d^3 {\bf{x}}e^{i{\bf{k.x}}}\frac{\delta\rho({\bf{x}})}{\rho_c}=&\int d z e^{i {\bk}_{||}.{z}}\delta(z)\frac{\sigma_w(\eta,\eta_i)}{\rho_c}  \tilde{F}(k_{\perp};\xi)\nonumber\\     
=&(2\pi) \frac{\sigma_w(\eta,\eta_i)}{\rho_c} \xi(\eta,\eta_i) \frac{J_1(k_{\perp}\xi(\eta,\eta_i))}{k_{\perp}}\,,
\end{align}
where $k_{\perp}=\sqrt{k^2-({\bf{k.\hat{n}}})^2}$. In order to simplify notation we use the notation $\overline{\sigma}_w(\eta,\eta_i)\equiv\sigma_w(\eta,\eta_i)/\rho_{c}$ in what follows. 

\section{Compensation}
\label{sec:compensation}

After topological defects are formed, there must be a compensating underdensity in the rest of the matter in the universe due to energy-momentum conservation.  This underdensity is not included in the particular solution generated by integrating the source with the Green's function, and so must be accounted for somehow. A procedure for incorporating energy-momentum conservation into the particular solution is called compensation  \cite{Albrecht:1991br}.


To understand this point in more detail, let us examine the $00$ component of the Einstein equation at linear order,
\ben
-k_ik_j\tilde{h}_{ij} + \frac{2}{3}k^2 h + 2\frac{\dot a}{a} \dot h = 6 \left( \frac{\dot a}{a}  \right)^2 \de_c + 16\pi G \Theta_{00},
\een
where $\tilde{h}_{ij}$ is the traceless part of the spatial metric perturbation and $h$ is its trace.  In the coordinate system which leads to (\ref{eq:linevol}), $\dot h = - 2 \dot \de_c$, and the equation can be rewritten 
\ben
-k_ik_j\tilde{h}_{ij} + \frac{2}{3}k^2 h  = 6 \left( \frac{\dot a}{a}  \right)^2 \de_c + 4\frac{\dot a}{a} \dot \de_c + 16\pi G \Theta_{00},
\een
Before the phase transition which produced the strings, the energy-momentum was uniform, and metric tensor was unperturbed.\footnote{At linear order the inflationary perturbations are independent and can be added separately.} After the transition, the metric tensor can be perturbed on scales inside the horizon, which means that $\tilde{h}_{ij}$ can grow no faster than white noise at low $k$. Hence, as $k \to 0$, 
\ben
\label{e:ComEqn}
6 \left(\frac{\dot a}{a}  \right)^2 \de_c + 4 \left(\frac{\dot a}{a}  \right) \dot \de_c + 16\pi G \Theta_{00} \to 0.
\een
The scaling property of the defect source requires that $\Theta_{00} \sim \eta^{-1/2}$ as $k\eta \to 0$, which means that $\delta_c \sim \eta^{3/2}$. 
However, 
when we calculate the particular solution of the equations~\eqref{eq:linevol} with the Green's function, 
\begin{align}\label{eq:greensmatter}
\delta_c^S({\mathbf{k}},\eta) &= \int_{\etaInit}^\eta d\eta'\mathcal{G}_c(k;\eta,\eta') \Theta_+({\mathbf{k}},\eta'),
\end{align}
we find that this ``subsequent'' perturbation inevitably contains a super-horizon growing mode proportional to $\eta^2$.  
The resolution, as pointed out by Albrecht and Stebbins \cite{Albrecht:1991br}, is that the growing mode is cancelled or ``compensated'' by the homogeneous part of the solution $\delta_c^I$, set at the initial time $\etaInit$.


In \cite{Albrecht:1991br,9712008} the effect of compensation was modelled by multiplying the synchronous gauge source  by a kernel 
\ben
\label{e:ComKer}
\ga_c(k,\eta) = \frac{1}{1 + k^2_c(\eta)/k^2},
\een
 where $k_c \sim \eta^{-1}$.  
This has the effect of multiplying $\de_c$ by a factor $(k\eta)^2$ for $k\eta \ll 1$, which eliminates the super-horizon growing mode, but allows perturbations to grow within the horizon where they are observable. This is a useful trick when one wishes to avoid generating spurious growing modes in numerical integration of the Green's function. 
In fact, any factor which removes the divergent term is acceptable, provided it is normalised to reproduce the correct amplitude of the growing mode inside the horizon.  
In \cite{Albrecht:1991br}, $\eta k_c(\eta)$ was chosen as $2\pi$ or $4\pi$, while Avelino et al used a function advocated by Cheung and Magueijo \cite{9702041} interpolating between $\sqrt{6}$ in the radiation era and $\sqrt{18}$ in the matter era.

Cheung and Magueijo's argument was given in the flat-slicing gauge, not the synchronous gauge used here. Nonetheless, we will see that using their choice of compensation kernel $k_c(\eta) = \sqrt{18}/\eta$ with our model for the source (described below) results in a power spectrum close to that produced by the commonly-used CMBACT  \cite{CMBACT} software, with appropriately chosen parameters. \cmbact\ integrates the synchronous equations directly, rather than using a Green's function, and does not produce a super-horizon growing mode.
Our attitude will be that the agreement is sufficient justification for using the combination of the source model and the compensation kernel we choose. 


\section{Comparison of matter power spectra}\label{sec:powerspecSEC}
\subsection{UETC approach}
\subsubsection{General Formalism}
The matter power spectrum, $P(k)$, is defined using the two point correlator of the perturbation, $\delta_c$, by the expression
\ben\label{e:PowSpeDef}
\langle \delta_c(\bk_1,\eta)\delta_c(\bk_2,\eta)\rangle = (2\pi)^3 \delta^3(\bk_1+\bk_2) P(k_1,\eta)\,. 
\een
We introduce the dimensionless unequal time correlator $C_+$ by 
\begin{align}\label{eq:cplusdef}
\langle\Theta_+({\mathbf{k}},\eta) \Theta_+^*({\mathbf{k'}},\eta') \rangle=(2\pi)^3 \delta^3({\mathbf{k-k'}})\frac{\phi_0^4}{\sqrt{\eta \eta'}}C_+(k,\eta,\eta') \,,
\end{align}
where $\phi_0$ is the expectation value of the symmetry-breaking field, and is related to the string tension by $\mu \simeq 2\pi \phi_0^2$ in the Abelian Higgs model.  A cosmic string network generically evolves towards a scaling regime, by which we mean that $C_+$ is a function of the dimensionless combinations $x=k\eta$ and $x'=k\eta'$ \cite{PenSpergelTurok,DurrerKunzMelchiorri}. It is also convenient to use the variables
\ben
z = \sqrt{k^2\eta\eta'}, \quad r = \eta'/\eta\,.
\een
As described in Section~\ref{sec:compensation} the source term in the equation of motion \eqref{eq:linevol} should be multiplied by the compensation factor $\gamma_c(k,\eta)$. Therefore, the perturbations are given by the altered form of equation~\eqref{deltaNsol}:
\ben
\delta_N(\bk,\eta)=\int_{\eta_i}^\eta \d\eta' \GG_N(k;\eta,\eta')\Th_+(\bk,\eta')\gamma_c(k,\eta')\,.
\een
The power spectrum of the matter density perturbation is then given by 
\begin{align}\label{e:SubPSdef}
P(k) = \ep^2  \int_{\eta_i}^{\eta}d\eta' \int_{\eta_i}^{\eta}d\eta'' {\GG_c(k;\eta,\eta')\GG_c^*(k;\eta,\eta'')} \frac{C_+(k\eta',k\eta'')}{\sqrt{\eta'\eta''}}\gamma_c(k,\eta')\gamma_c(k,\eta'')\,.
\end{align}
where we set $\ep\equiv 4\pi G\phi_0^2$ to compress notation.
The UETC is expected to be well approximated by the generic form \cite{DurrerKunz,Bevis06,Bevis10}
\ben
\label{e:GenForUETC}
C_+(z,r) = \left\{
\ba{lc}
\displaystyle \frac{2\bar{E}_+}{r^\frac{3}{2} + r^{-\frac{3}{2}}} & z < 1 \\[12pt]
\displaystyle E_+(z)e^{-z^2\ln(r)^2/2A^2} & z \ge 1 
\ea
\right.
\een
where $A$ is constant and $E_+(z)$ approaches a constant value, $\bar{E}_+$, on large scales (for $z\ll1$) \cite{Bevis10}. They generally take different values in the radiation (r) and matter (m) eras, with $E_+^\text{r} > E_+^\text{m}$. 
The appearance of the factor $2/(r^{3/2}+r^{-3/2})$ on large scales is explained in  Appendix \ref{sec:AppA}, and $z^2\ln(r)^2$ is a useful approximation to  $k^2(\et - \et')^2$, observed in numerical simulations \cite{Bevis10}.

Combining the large scale and small scale behaviour leads to the more convenient form of the UETC,
\ben\label{eq:cplusanalytic}
C_+(k,\eta,\eta')=\frac{2 E_+(z)}{(r^\frac{3}{2} + r^{-\frac{3}{2}})} e^{-z^2\ln(r)^2/2A^2}\,.
\een

%
%
\subsubsection{Gaussian model for the cosmic string \uetc}\label{subsubsec:gaussian}
Using the expression for $\Theta_+$ in equation~\eqref{eq:thetaplus} for a cosmic string network (evolving according to a Nambu-Goto action) we may write
\begin{align}\label{eq:thetaplusproduct}
\langle\Theta_+({\mathbf{k}},\eta) \Theta_+^*({\mathbf{k'}},\eta') \rangle=
(2\mu)^2 \int ds ds'  \vev{{\dot\bXs_s^2} {\dot\bXs_{s'}^2}  e^{i(\mathbf{k \cdot \Xs_s} -\mathbf{k' \cdot \bXs_{s'})} } }\,,
\end{align}
where the subscript $s'$ indicates that we take $\bXs(\eta',s')$. 

For modes well inside the horizon, we approximate the string network as consisting of an ensemble of randomly placed strings with a Gaussian distribution for the random fields $\dot\bXs$, $\bXs'$, as in \cite{Hind93,HRS09x1}.

The important correlation functions are denoted:
\bea
\Gamma(s_-,\eta) &=& \langle [\bXs(s,\eta)- \bXs(s',\eta)]^2\rangle\,, \qquad
\Pi(s_-,\eta) = \langle (\bXs(s,\eta)- \bXs(s',\eta))\cdot\dot{\bXs}(s',\eta)\rangle\,, \nonumber\\
V(s_-,\eta) &=& \langle \dot{\bXs}(s,\eta) \cdot \dot{\bXs}(s',\eta)\rangle\,,
\eea
where $s_-\equiv s-s'$. We will also write $\eta_- = \eta - \eta'$, $s_+ = (s+s')/2$ and $\eta_+ = (\eta+\eta')/2$.
For small time differences and small spatial separation between points on the string (such that $|\eta_-|<\xi$ and $|s_-|<\xi$), we can approximate the two point function $\langle [\bXs(\sigma,\eta)- \bXs(\sigma',\eta')]^2 \rangle$ in the following form,
\begin{align}
\langle [\bXs(s,\eta)- \bXs(s',\eta')]^2 \rangle&\approx \langle [\bXs(s,\eta_+)- \bXs(s',\eta_+)- \dot{\bXs}(s',\eta_+)(\eta_-)]^2 \rangle \nonumber\\
&\approx \Gamma(s_-,\eta_+)+\Pi(s_-,\eta_+)\eta_-+V(0,\eta_+) \eta_-^2\,,
\end{align}
The asymptotic small scale limit for these functions is given by 
\ben
\Gamma(s)\approx \overline{t}^2s^2,\,\Pi(s)\approx c_0 s/(2\xi),\,V(s)\approx \overline{v}^2,
\een 
where the correlation length $\xi$ at time $\eta$ obeys $\xi\propto \eta$. 
The constant $c_0$ is small thanks to the approximate time-reversal symmetry of the string network\footnote{Wandelt private communication} \cite{HRS09x1} and terms involving $\Pi$ will therefore be neglected in the remainder of the paper. 
Note the constraint condition $\vba^2 + \tba^2 = 1$.

For larger scales we assume the correlators between $\dot\bXs$ and $\bXs'$ vanish, as is appropriate for a random walk. A suitable model for the velocity correlator is \cite{Hind93,ReganShellard09}
\begin{align}\label{Vdefin}
V(s_-,\eta)\approx \overline{v}^2\left( 1-\frac{|s_-|}{\xi}\right) \exp(-|s_-|/\xi).
\end{align} 
However, using this function leads to complicated expressions for the UETC so instead we simplify and set\footnote{The suppression due to the exponential function ensures that this approximation is reasonable.} $V(s_-,\eta)\approx \vba^2/2$. 
It is also useful to think of the string network as being made up of independent segments of length $\xi$, in which case we should limit the range of integration over $s_-$ to $(-\xi,\xi)$.  There exists a certain ambiguity regarding the time at which the correlation length - which is assumed to be equal to the inter-string distance, - is evaluated. A sensible choice may be to set $\xi ={\rm min} (\xi(\eta),\xi(\eta'))$, since the correlation length scales as $\xi(\eta) \propto \eta$, and in computing $\int ds ds'$ in equation \eqref{eq:thetaplusproduct}, the Gaussian approximation for the string correlation functions is a better approximation at the smallest available length-scale. Alternatively, given that the UETC must be symmetric under the exchange $\eta\leftrightarrow \eta'$ one may express $\xi$ in the symmetric form $\xi=\sqrt{\xi(\eta),\xi(\eta')}$. In practice, the choice makes little difference, since on sub horizon scales, the UETC peaks strongly in the equal time limit, while we will account for the superhorizon behaviour explicitly.
We make use of the latter identification in this section and set $\xi= \alpha\sqrt{\eta\eta'}$, where $\al \simeq 0.15$ is a constant. 

Using the relationships described here we obtain the following form for the UETC $C_+$
\begin{align}
C_+(k\eta,k\eta') &=4\bar\mu^2 \frac{\sqrt{\eta\eta'}}{\Om}\int ds_+ ds_- \left(V(0)^2 +\frac{2}{3} V(s_-)^2\right)\exp\left(\frac{-k^2\Gamma^2(s_-)}{6}\right)\exp\left(\frac{-k^2\overline{v}^2\eta_-^2}{6}\right)\,,\nonumber
\end{align}
where $\Om$ is a normalisation volume, and $\bar\mu = \mu/\phi_0^2$. 
Therefore, we find
\begin{align}\label{eq:F2eqn}
C_+(k\eta,k\eta')
&\approx 4\bar\mu^2\sqrt{\eta\eta'}\Ll\int_{-\xi}^{\xi} ds_- \frac{7}{6}\overline{v}^4\exp\left(-\frac{k^2\overline{t}^2s_-^2}{6}\right)\exp\left(\frac{-k^2\overline{v}^2\eta_-^2}{6}\right),
\end{align}
where $\Ll = \int ds_+/\Om$.
Cosmic strings obey a scaling regime such that the energy density of long strings is given by $\rho_{\infty}=\mu/\xi^2$. This implies that $\Ll=1/\xi^2$. Hence, correcting for the super-horizon behaviour as in equation~\eqref{eq:cplusanalytic}, we obtain
\begin{align}\label{e:StrModCor}
C_+(k\eta,k\eta')
&\approx 4\bar\mu^2 \frac{7}{6}\frac{\sqrt{6\pi}\vba^4}{\al^2\overline{t}}\frac{1}{k\sqrt{\eta\eta'}}{\rm{erf}}\left(k\overline{t}\al\sqrt{\frac{{\et\et'}}{{6}}} \right)
\exp\left(\frac{-k^2\overline{v}^2\eta_-^2}{6}\right) \frac{2}{r^{3/2}+r^{-3/2}}\,,
\end{align}
where $r=\eta/\eta'$. Comparison with the general UETC form (\ref{e:GenForUETC}) shows that 
\ben
\label{e:EplusAvals}
E_+(z) = \frac{14\sqrt{6\pi}}{3}\frac{\bar\mu^2\vba^4\ga_\vba}{\al^2 z}{\rm{erf}}\left(\frac{\al z}{\sqrt{6}\ga_\vba} \right), \quad A = \frac{\sqrt{3}}{\vba},
\een
where we have used $\tba = \ga_\vba^{-1}$.
Note the characteristic stringy $z^{-1}$ behaviour as $z \to \infty$.
The approximate values of the parameters for Abelian Higgs and Nambu-Goto strings are given in Table \ref{t:pars}. 
The values of the parameters are constant in the deep radiation or deep matter era. In order to track their evolution in the matter-radiation transition we make use of the velocity-one scale (VOS) model \cite{Kibble1985,9307325,9602271,0003298}, by evolving the equations
\bea
\frac{d\ln \xi}{d\eta}=(1+\vba^2)\frac{d\ln a}{d\eta} + \frac{\tilde{c} \vba}{2 \xi}\,,\qquad
\frac{d\vba}{d\eta}=(1-\vba^2)\left(\frac{\tilde{k}}{\xi}-2 \frac{d\ln a}{d\eta}\vba\right)\,,
\eea
where $\tilde{c}$ is a constant quantifying the energy loss rate (set to a value $0.23$ to agree with CMBACT), and 
$
\tilde{k}=({2\sqrt{2}}/{\pi}){(1-8\vba^6)}/{(1+8\vba^6} )
$
\cite{0003298}.

By substituting the expression for the UETC given by equation~\eqref{e:StrModCor} into equation~\eqref{e:SubPSdef} one obtains our formula for the matter power spectrum.

\begin{table}[ht]
\centering
\renewcommand{\arraystretch}{1.2}
\begin{tabular}{|c|c|c|c|c|}
\hline
& \multicolumn{2}{|c|}{matter era} & \multicolumn{2}{|c|}{radiation era} \\
\hline
& AH & NG & AH & NG \\
\hline
$\bar\mu$ & $2\pi$ & $2\pi$ & $2\pi$ & $2\pi$ \\
$\vba$ & 0.51 & 0.59 & 0.5 & 0.63 \\
$\bar t$ & 0.86  & 0.81  & 0.87  & 0.77\\
$\alpha$ & 0.30 & 0.17 & 0.30 & 0.15\\
\hline
$\bar{E}_+$ &700 & 4140 & 640 & 7300\\
$A$ & 3.5 & 2.9  & 3.5 & 2.7  \\
$\sqrt{2\pi} A \bar{E}_+/\bar\mu^2$ & $ 45 $ & $ 260 $  & $ 40 $ & $ 460 $  \\
\hline
\end{tabular}
\caption{\label{t:pars}
Approximate values of cosmic string parameters in the Abelian Higgs \cite{Bevis10,08121929} and Nambu-Goto \cite{11015173} cosmic string scenarios, with the derived value of the UETC parameters $E_+$ and $A$ defined in (\ref{e:GenForUETC}).}
\end{table}

\subsubsection{Comparison to USM}

CMBACT is based on the unconnected segment model (USM) model \cite{9711121,Pogosian:1999np} where the string network is modelled as a set of randomly placed and oriented straight segments moving with speed $\bar{v}$ in a random direction at right angles to the orientation. 
Scaling demands that the length of the segments grows with time, and their density decreases as the inverse square.  

\begin{figure}
\centering
\vspace{0.25cm}
\hspace{0.1cm}
{\includegraphics[width=0.44\linewidth]{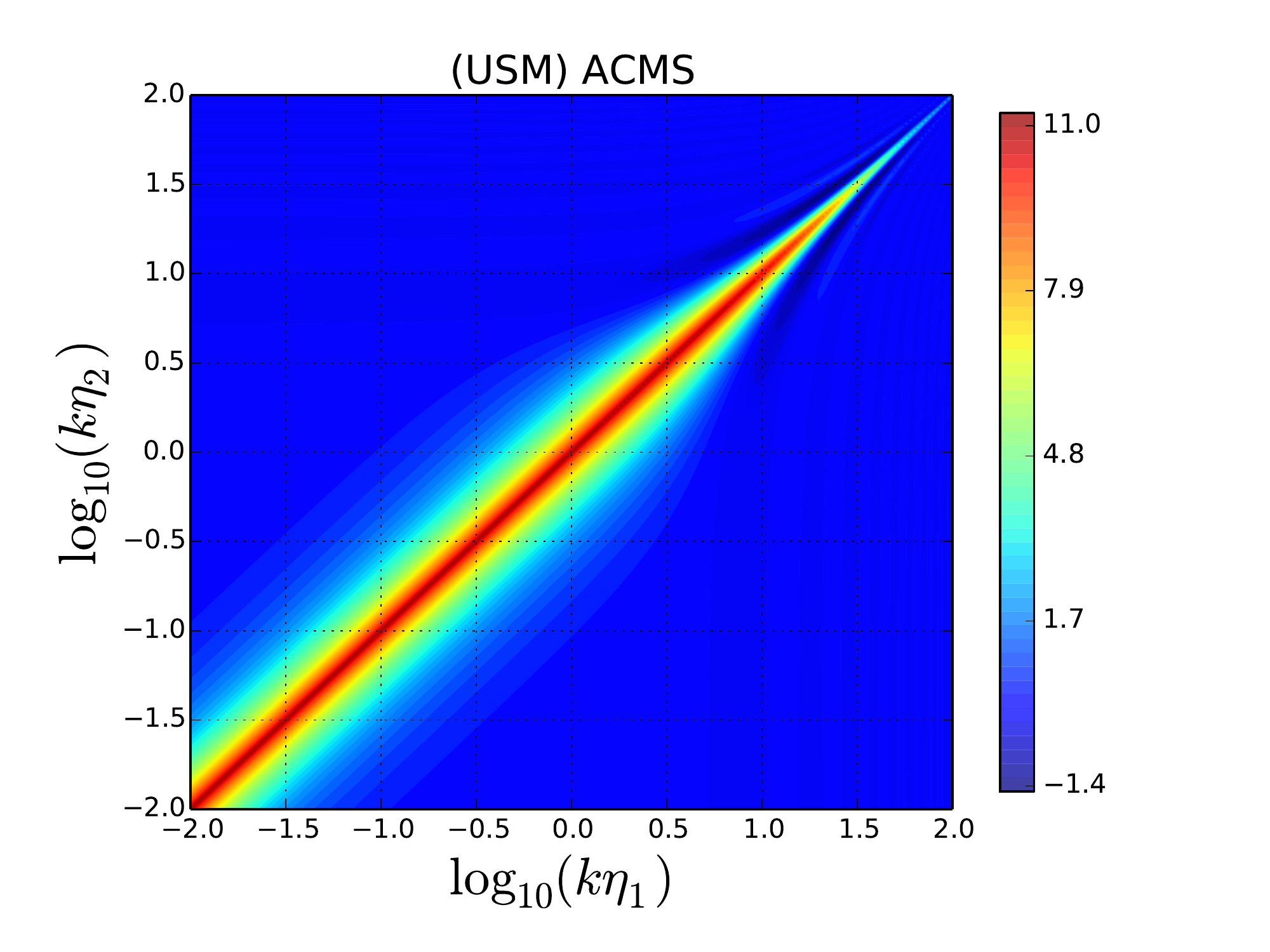}}
\hspace{0cm}
{\includegraphics[width=0.44\linewidth]{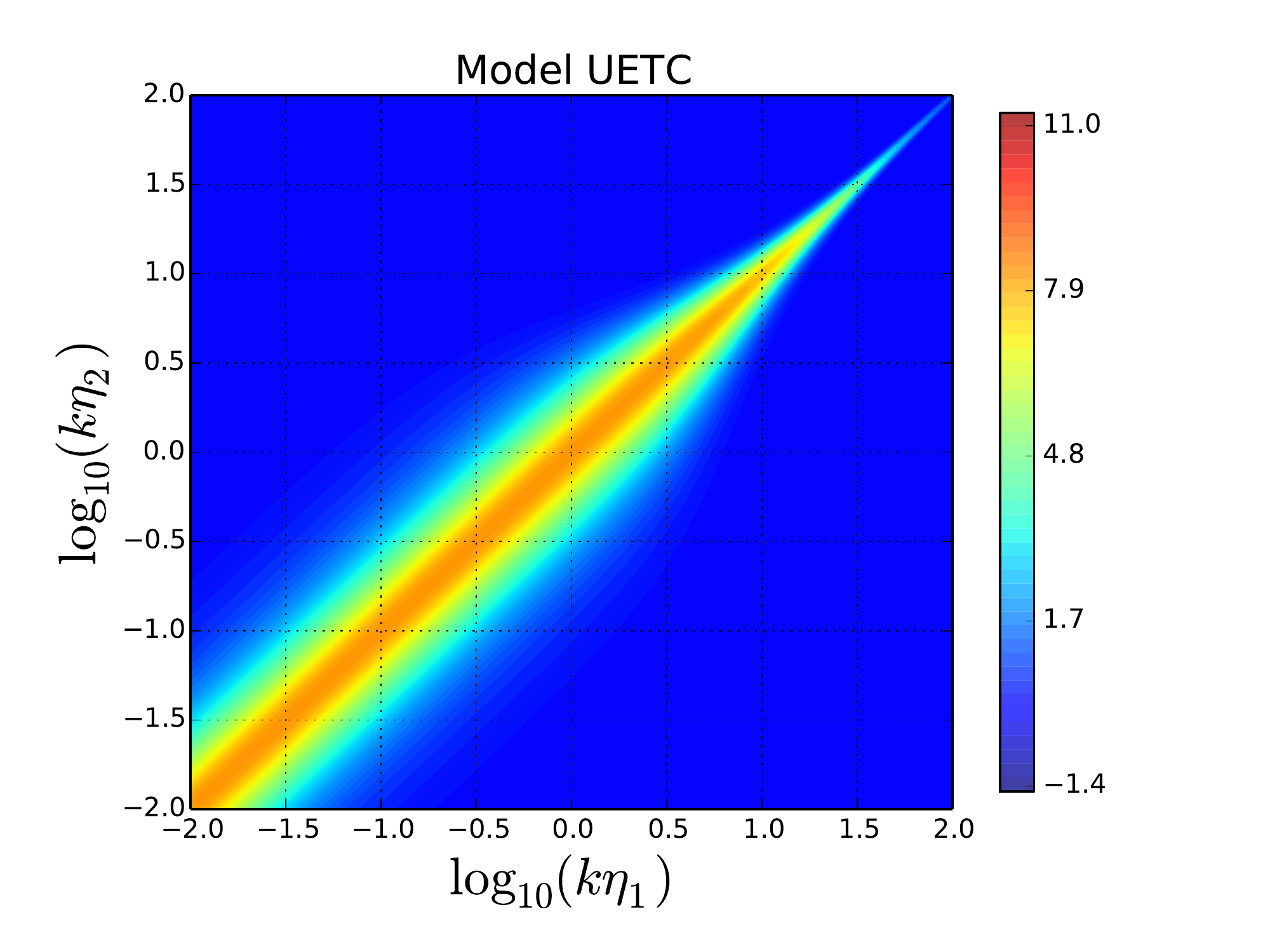}}\\
{\includegraphics[width=0.44\linewidth]{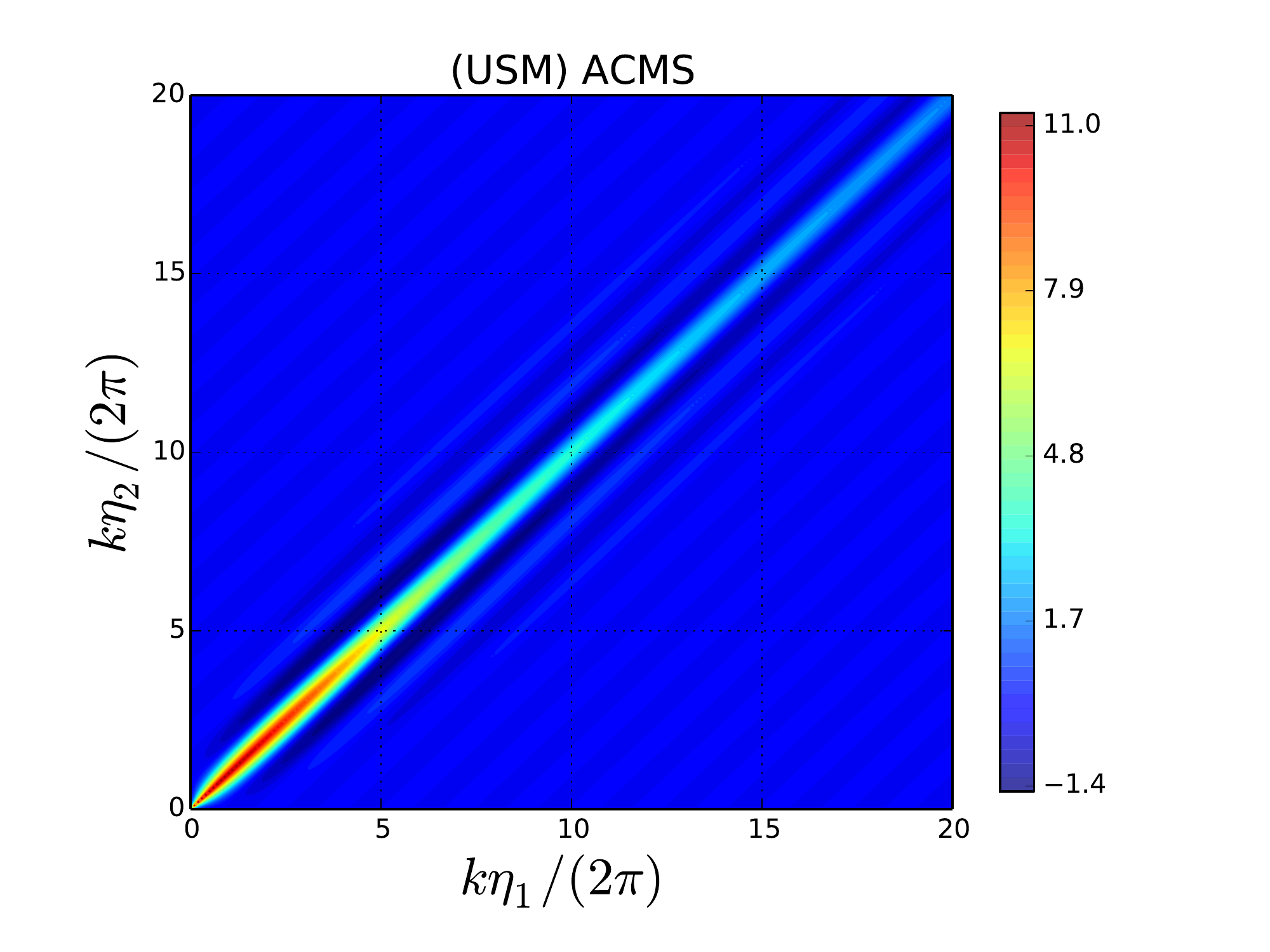}} 
{\includegraphics[width=0.44\linewidth]{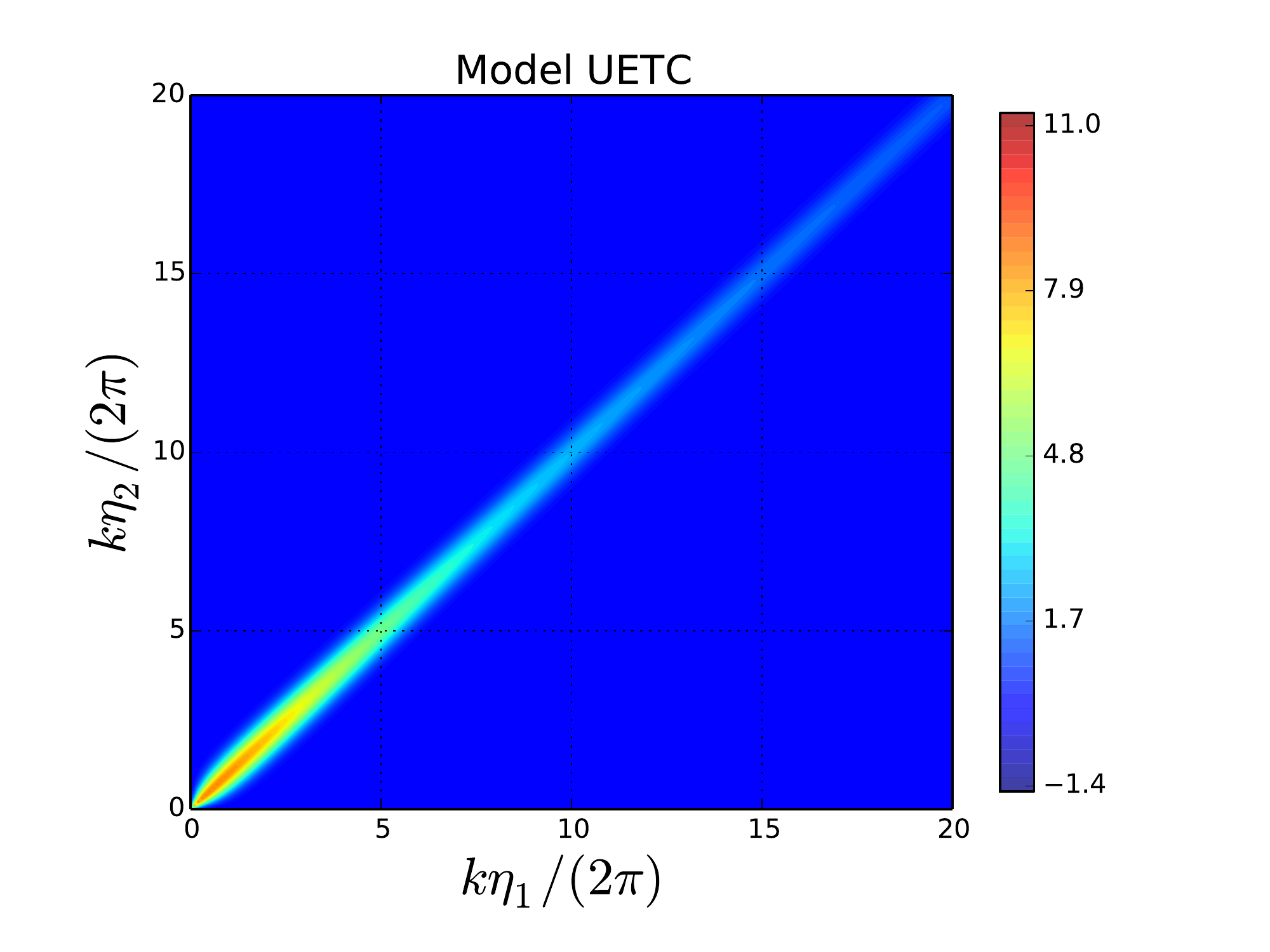}} 

\caption{Comparison of the unequal time correlator computed using the USM model (and, in particular, the analytic version given by ACSM \cite{12092461}) with that computed in this paper. As is clear from the lower panel the two approaches give very similar qualitative behaviour. The upper panel, on a log scale, highlights the slight difference between the two, with the UETC computed here having a broader spread around the equal time result. The approach in this paper ensures the UETC is everywhere non-negative, while the UETC computed using the ACSM (USM) model contains regions of negative magnitude.}
\label{fig:UETCs}
\end{figure}

As we shall show in this section, the UETC of the USM and that computed in the Gaussian string model are very similar.  
In \cite{12092461} (hereafter referred to as ACMS) analytic expressions were derived for the UETCs of the USM. 
Although the UETC of  $\Theta_+$ was not directly computed, we can use 
$\langle\Th_+ \Th_+^*\rangle \simeq 4\vba^4 \langle\Th_{00} \Th_{00}^*\rangle$. We will also assume that the effective string tension and string energy per unit length are equal, as in the current version of CMBACT, and that string segments decay instantaneously.
Hence, in the same notation as ACMS, 
\ben\label{eq:uetcUSM}
\langle\Th_+(\bk,\eta_1) \Th_+^*(\bk,\eta_2)\rangle \simeq \frac{8 \vba^4 \mu^2}{k^2 (1-\vba^2)} f(\eta_1,\eta_2,\xi,1) [I_1(x_-,\rho)-I_1(x_+,\rho)]\,,
\een
where $x_{1,2}=k\alpha \eta_{1,2}$, $x_{\pm} = (x_1\pm x_2)/2$, $\rho=k |\eta_1-\eta_2|\vba$. The functions are 
\ben
I_1(x,\rho)=\sum_{c=0}^{\infty}\frac{1}{c!}\frac{\rho}{2c-1}\left(\frac{-x^2}{2\rho}\right)^c j_{c-1}(\rho)\,,
\een
where $j_n(x)$ is the spherical Bessel function, and  
\ben
f(\eta_1,\eta_2,\xi,1) = \frac{1}{\alpha^3 {\rm Max}(\eta_1,\eta_2)^3}, 
\een
In Figure~\ref{fig:UETCs} we plot a comparison of the analytic expression for the UETC calculated using the USM (\ref{eq:uetcUSM}) with the UETC computed using the Gaussian model in this paper (\ref{eq:cplusdef},\ref{e:StrModCor}). 
The qualitative behaviour is very similar. The correlator peaks at equal times, and decreases as a function of the ratio $x_1/x_2$ outside the horizon, or $|x_1 - x_2|$ inside the horizon. 
Although it is not obvious, the super horizon behaviour at large time ratios is very close to each other, as in the ACMS expression, 
\ben
\langle\Th_+(\bk,\eta_1) \Th_+^*(\bk,\eta_2)\rangle \to
\frac{8 \vba^4 \mu^2}{\alpha^3 (1-\vba^2)}
 \frac{1}{(\eta_1 \eta_2)^{1/2}}\frac{1}{{\rm Max}(r^{3/2},r^{-3/2})}\,
\een
where $r=\eta_1/\eta_2$.
This is to be compared to the large $r$, small $z$ behaviour of Eq.~\eqref{eq:cplusanalytic}), which implies 
\ben
\langle\Th_+(\bk,\eta_1) \Th_+^*(\bk,\eta_2)\rangle \to
\frac{14}{3}\sqrt{\frac{\pi}{6}}\frac{\vba^4\mu^2}{\al} \frac{1}{(\eta_1 \eta_2)^{1/2}} \frac{2}{r^{3/2}+r^{-3/2}}
\een 
It should also be noted that the \uetc\ can take small negative values in the ACMS expression, possibly due to truncation errors in the computation of the series expansion, with their magnitude small enough that the question of their sign is irrelevant; by contrast the UETC using the Gaussian model is positive everywhere. The Gaussian model UETC has smaller amplitude but has a broader peak than the ACMS version; as we shall observe, the resulting power spectrum using either UETC expression will have a similar amplitude.

 
\subsection{Matter power spectrum in the Wake model}\label{sec:wakemodelpower}
	In order to compute the power spectrum using the wake model density perturbation, \eqref{eq:delta1}, we first compute the average over all orientations $\hat{\bf{n}}$ of the square of the density perturbation from a single wake. Setting $\hat{\bf{k}}\cdot\hat{\bf{n}}=\mu$, we find that this quantity is given by
\begin{align}
\langle |\delta_1(\bk,\eta,\eta_i)|^2\rangle_{\bf{\hat{n}}}&=\int \frac{d\hat{\bn}}{4\pi} |\delta_1({\bk,\eta,\eta_i)}|^2= \left(2\pi{\bar{\sigma}_w(\eta,\eta_i)} \xi(\eta_i)^2\right)^2 \int_{-1}^1 \frac{d\mu}{2} \frac{\left[J_1(k\xi(\eta_i)\sqrt{1-\mu^2})\right]^2}{k^2\xi(\eta_i)^2 (1-\mu^2)}\,.
\end{align}
The total angle averaged (squared) density perturbation at time $\eta$ due to all wakes formed at time $\eta_i$ is thus given by
$
dP(k,\eta,\eta_i) = d\nwake(\eta_i){\langle |\delta_1(\bk,\eta,\eta_i)|^2 \rangle_{\bf{\hat{n}}}}\,,
$
where $d\nwake(\eta_i)$ is the number density of wakes formed between $\eta_i$ and $\eta_i + d\eta_i$. 
Using the scaling assumption, the number density of wakes formed in time $d\eta_i$ can depend only on the time of formation, and so  
\ben
{d\nwake}=\frac{\nu}{\eta_i^4}d\eta_i,
\een 
where $\nu$ is a constant. 
Averaging over the angles to compute the power spectrum implicitly results in an isotropic power spectrum, such that the result is independent of configuration. We note however, that we have neglected the issue of compensation terms in this expression. Since the wake model utilises a Zel'dovich approximation which should agree on linear scales with the result computed using perturbation theory, it is clear that such terms are necessary and must be incorporated. Since the effect of compensation are accounted for in linear perturbation theory by multiplying the source by the factor $\gamma_c$ - as given in Section~\ref{sec:compensation} - we infer that we may approximate the effect of compensations in the wake model by the replacement
\ben
\delta_1(\bk,\eta,\eta_i) \rightarrow \delta_1(\bk,\eta,\eta_i)  \gamma_c(k,\eta_i)\,,
\een 
i.e. by multiplying the expression for the density by the compensation factor for the time the wake was sourced.

We infer that the total power spectrum, defined in (\ref{e:PowSpeDef}), is given at time $\eta$ by
\begin{align}\label{eq:powerwake}
P(k,\eta)&=\int_{\eta_1}^{\eta} d\eta_i \frac{d\nwake}{d\eta_i}   {\langle |\delta_1(\bk,\eta,\eta_i)|^2 \rangle_{\bf{\hat{n}}}} \gamma_c(k,\eta_i)^2\nonumber\\
&=\int_{\eta_1}^\eta d\eta_i \left[\frac{\nu}{\eta_i^4}\left(2\pi\xi(\eta_i)^2{{\bar{\sigma}_w}(\eta,\eta_i) \gamma_c(k,\eta_i)}{} \right)^2 \int_{-1}^1 \frac{d\mu}{2} \frac{\left[J_1(k\xi(\eta_i)\sqrt{1-\mu^2})\right]^2}{k^2\xi(\eta_i)^2 (1-\mu^2)}\right]\,,
\end{align}
where $\eta_1$ is an arbitrary initial time\footnote{The wake model makes use of the Zel'dovich approximation which is valid only in the matter era. Therefore, we must simply require $\eta_1\gtrsim \eta_{\rm{eq}}$.}. 
We can relate the constant $\nu$ to the string network parameters as follows:  the area of new wakes generated in time $d\eta_i$ is $\pi(\al\eta_i)^2 d\nwake$, which is equal to the area swept out by the string network in the same time interval, or
\ben
\pi(\al\eta_i)^2{d\nwake} =  {\cal L} \tba \vba d\eta_i,
\een
where the factor of the RMS tangent vector $\bar{t}$ translates from the invariant length density $\cL$ to the required length density. Hence, recalling that $\cL = 1/\al^2\eta_i^2$, we have
\ben
\label{e:nuDef}
\nu = \frac{\bar{t}\vba}{\pi^2\al^4}\,.
\een
The wake model is valid only for string sources deep within the matter era. Therefore, while we do not expect this approach to give highly accurate quantitative results, we expect the behaviour to be qualitatively similar to that computed using the linear approach. In particular our motivation for introducing the wake model is to validate the qualitative behaviour of the power spectrum and bispectrum computed using the linear model. In order to investigate the qualitative behaviour for the power spectrum, we shall compare the behaviour to that of the linear model in the deep matter era.

\subsection{Analytic limits and numerical evaluation}


\begin{figure}
\centering
\hspace{0.1cm}
{\includegraphics[width=0.44\linewidth]{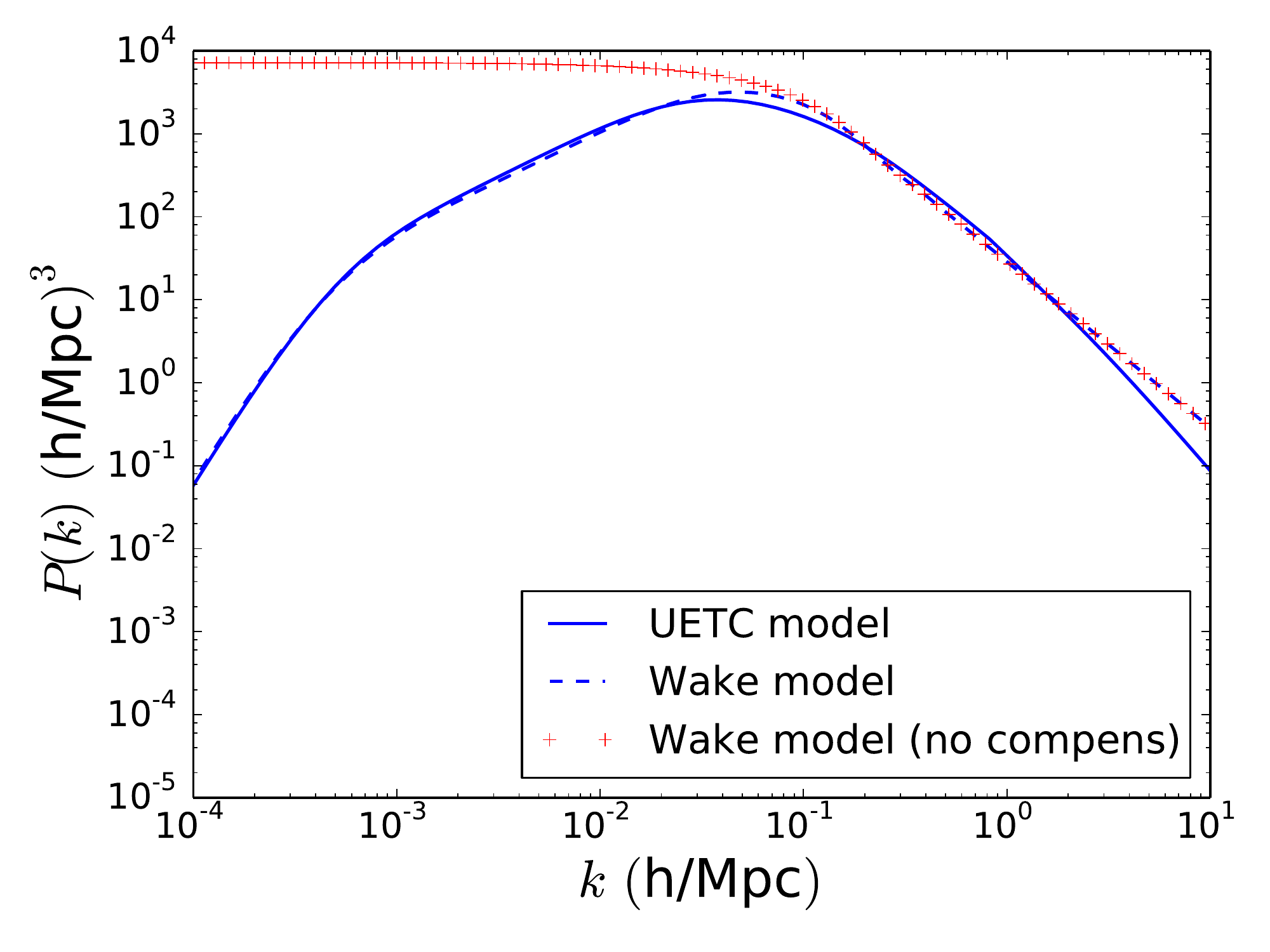}}
\caption{Comparison of the power spectrum computed for the wake model and the UETC (Gaussian) model in the matter dominated era. Also plotted is the wake model power spectrum without the compensation factor.}
\label{fig:wakemodelplot}
\end{figure}

\subsubsection{Analytic comparison of the wake model and the Gaussian UETC model }
In Appendix~\ref{sec:AppC} we carry out some analytic estimates for both models. The wake model is strictly valid only in the matter dominated era, and in order to make comparisons to the Gaussian UETC model, we restrict the latter to the same era, such that the Green's function may be approximated as $\GG_c(k'\eta,\eta')=\eta^2/(5\eta')$. From the analytic results, we find that the wake model and Gaussian UETC model give qualitatively similar results with both describing a power spectrum proportional to $k^4$ on superhorizon scales, $k$ on large scales, and $k^{-2}$ on small scales, respectively. This agreement gives a reassurance that, despite the inherent non-linearity of a cosmic string network, the linear perturbation theory result produces accurate results. We shall carry out a similar analysis for the bispectrum. In order to investigate any quantitative differences between the two models, we numerically integrate equations~\eqref{eq:powerwake} and~\eqref{e:SubPSdef} (the latter using the aforementioned Green's function and the unequal time correlator of the Gaussian model), from $\eta_{\rm eq}$ to $\eta$. We use fixed values for the parameters given in Table~\ref{t:pars}. This comparison is plotted in Figure~\ref{fig:wakemodelplot}. We observe that the agreement is not just qualitative, but quantitative as well. We also plot the power spectrum of the wake model without the compensation factors. We see for $k\lesssim 0.1 h/{\rm Mpc}$ that, without the compensation factor, we would infer a much larger amplitude for the power spectrum.

\subsubsection{Numerical evaluation}
\begin{figure}
\centering
\hspace{0.1cm}
{\includegraphics[width=0.44\linewidth]{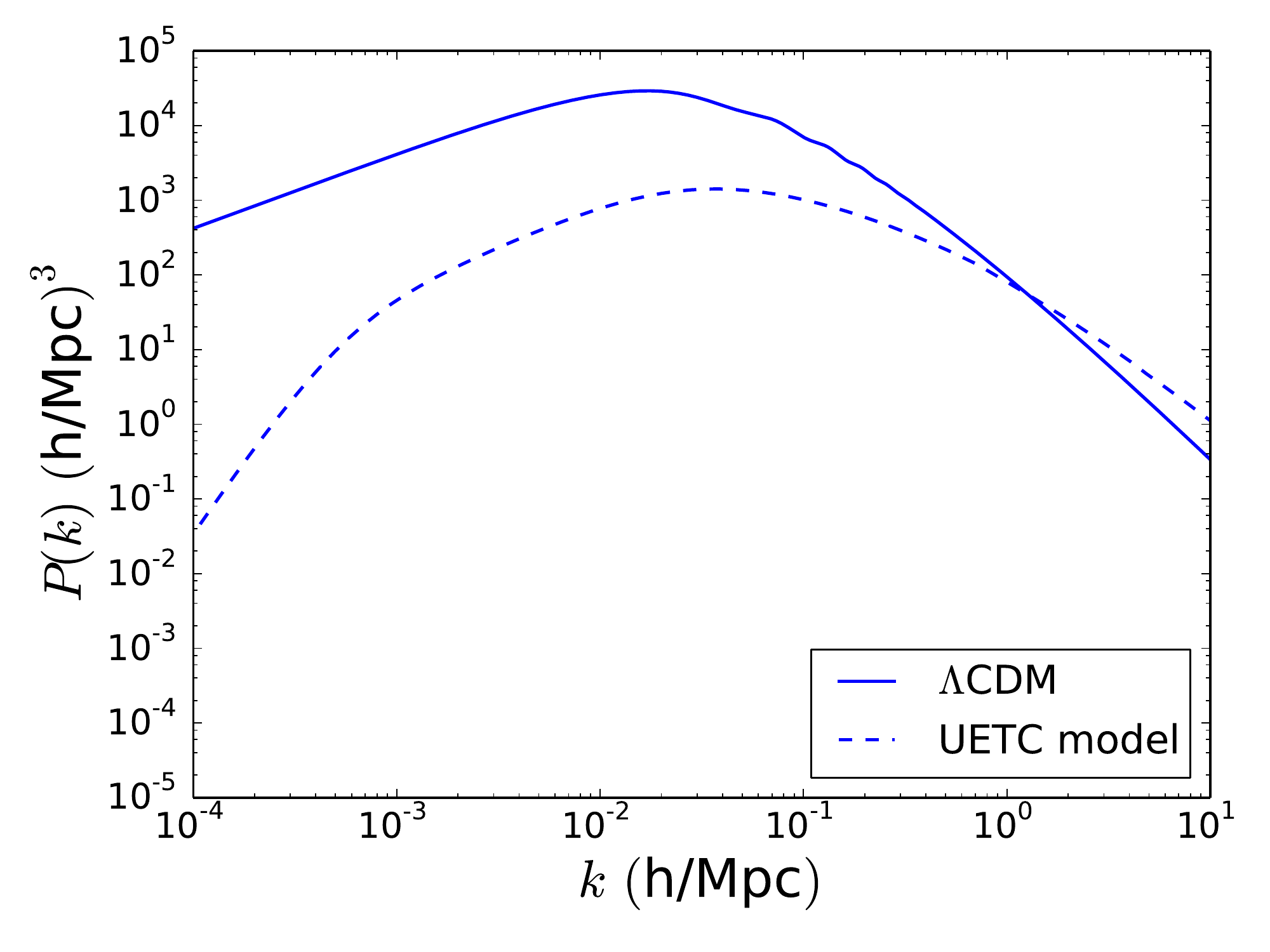}}
\hspace{0cm}
\vspace{0cm}
{\includegraphics[width=0.44\linewidth]{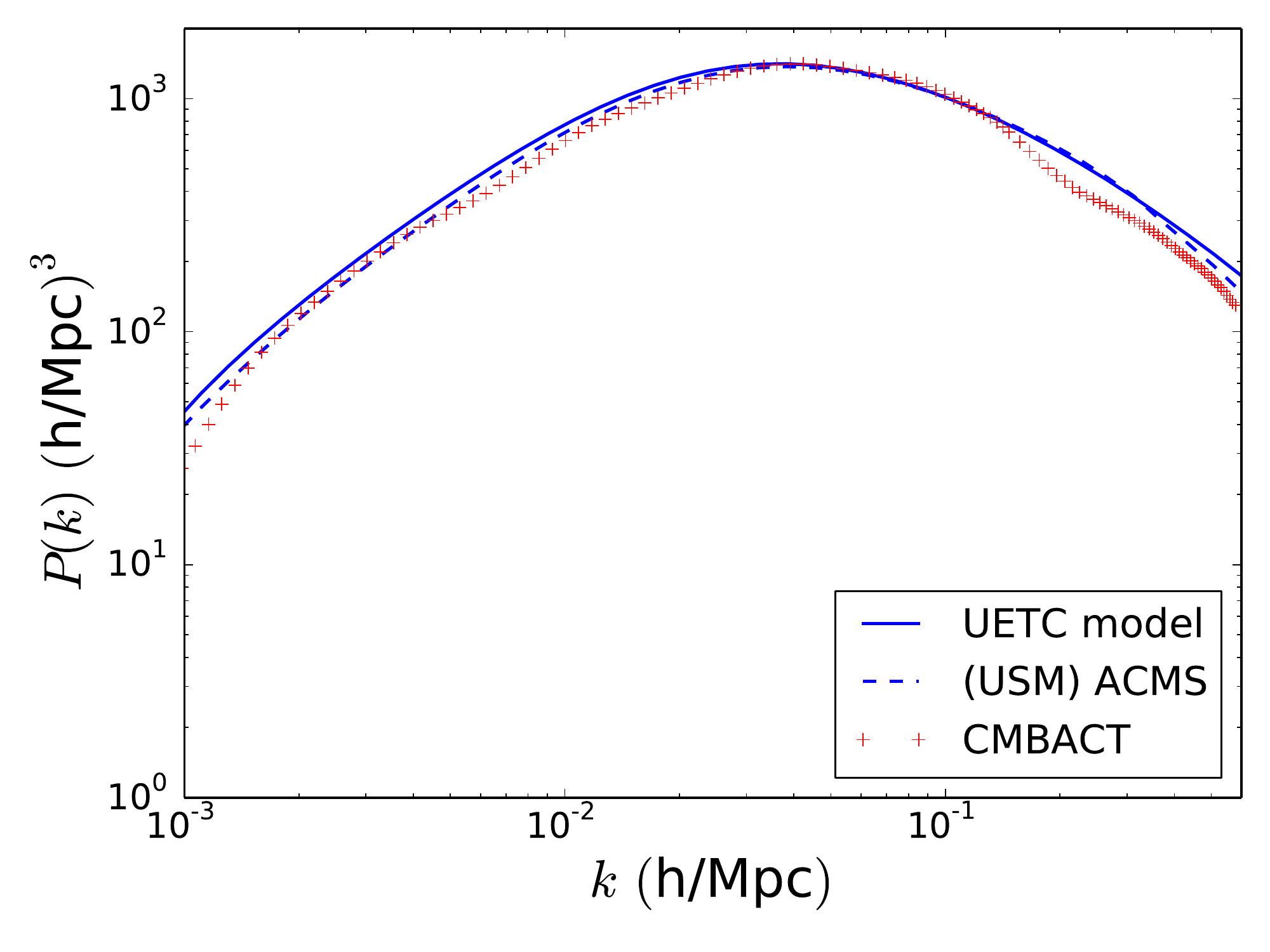}}
\caption{In the left panel we compare the inflationary perturbation induced matter power spectrum to that computed in this paper for cosmic strings. In the right panel we include a comparison of the matter power spectrum computed with the Green's function using (solid line) the ACMS computation for the UETC with that evaluated using (dash line) the Gaussian model UETC of this paper. We also include a comparison to the output from the CMBACT code using $200$ network realisations to calculate the UETCs. The relative lack of smoothness of the CMBACT power spectrum is due to the relatively low number of simulations used in its evaluation. The figures are computed for $z=0$ and for $G\mu=1.1\times 10^{-6}$. 
}
\label{fig:various1}
\end{figure}

We may now evaluate the power spectrum using the unequal time correlator using equation~\eqref{e:SubPSdef}. The Green's functions are computed numerically and the expressions integrated from $\eta_i=0.01$ to $\eta=14000$. The string parameters are evolved using the VOS model - described in Section~\ref{subsubsec:gaussian} - with initial conditions given by the radiation era scaling values in Table~\ref{t:pars} with the fiducial value of $G\mu$ given (as in CMBACT) by $G\mu=1.1\times 10^{-6}$. We also compute the power spectrum for the USM model by substituting equation~\eqref{eq:uetcUSM} into equation~\eqref{e:SubPSdef} (using equation~\eqref{eq:cplusdef}). The results are plotted in the right hand panel of Figure~\ref{fig:various1}. It is apparent that any difference between the two models are entirely trivial at the level of the power spectrum.
We also plot the comparison between these results and that obtained numerically with CMBACT using $200$ network realisations of the USM model\footnote{In this work we have not considered the presence of baryons, which must be present to run CMBACT. Therefore, for the running of CMBACT we have set $\Omega_m=0.25$, $\Omega_b=0.02$, and scaled the output by $(0.27/\Omega_m)^2$, to correct the amplitude.}. The comparison establishes that the compensation factor allows for the power spectrum to be correctly calculated using the Green's function technique.
 In addition we plot (in the left panel) the comparison
between the (linear) matter power spectrum for the matter component of the $\Lambda {\rm CDM}$ model, as computed using CAMB (with parameters given by results from the Planck satellite \cite{Ade:2013zuv}), emphasising that in the linear regime ($k\lesssim 0.3 h/Mpc$) that the cosmic string matter power spectrum lies at least an order of magnitude below that of the standard $\Lambda {\rm CDM}$ structure formation scenario (noting also that $P(k)\propto (G\mu)^2$). 

For these plots we have considered the spectra today, i.e. at  $\eta\approx 14000 {\rm Mpc}$. One may wonder whether the cosmic string spectrum may dominate at higher redshifts. In the matter era the scale factor has the conformal time behaviour $a\propto \eta^2$. Therefore, we infer from the analytic expressions in this section that $P(k)\propto 1/(1+z)^2$. Perturbations from inflation, $\Phi$, result in the density perturbation, $\delta$, which may be inferred from the Poisson equation
\ben
\delta(\bk,z)=\frac{2}{3}\frac{k^2 T(k) D(z)}{\Omega_m H_0^2}\Phi(\bk)\equiv M(k,z)\Phi(\bk)\,,
\een
where the linear transfer function, $T(k)$, is normalised to unity at large scales, and $D(z)$ is the linear growth function and is normalised $D(z)=1/(1+z)$ during matter domination. Therefore, we infer that the $\Lambda {\rm CDM}$ inflation based matter power spectrum, $\tilde{P}(k)$ also scales as $\tilde{P}(k)\propto 1/(1+z)^2$. Thus, we deduce that the cosmic string matter power spectrum at the current upper bound $G \mu\lesssim 1.3 \times 10^{-7}$ is around $3$ orders of magnitude below the $\Lambda {\rm CDM}$ linear matter power spectrum at all redshifts, and therefore offers little promise as a probe for cosmic strings.

\section{Bispectra}\label{sec:bispecSEC}

We recall that the matter bispectrum $B$ is given by the three point correlator of the matter perturbation,
\ben\label{eq:bispdef}
\langle \delta_c({\mathbf{k}_1 },\eta)\delta_c({\mathbf{k}_2 },\eta) \delta_c({\mathbf{k}_3 },\eta)  \rangle= (2\pi)^3 \delta^3(\bk_1+\bk_2+\bk_3) B(k_1,k_2,k_3,\eta)\,.
\een
In this section we describe our computations of the matter bispectrum due to the presence of a network of cosmic strings. We calculate $B$ in two different ways: in linear perturbation theory and in the wake model.

In linear perturbation theory, we need to evaluate the three-point unequal time correlator (\uetc3) of $\Theta_+$, which is prohibitively expensive in field theory simulations and in the USM framework. However, the Gaussian model gives a rather simple form,  and integrating it with the relevant Green's function is numerically tractable, and also allows for analytic estimates of the bispectrum in various regimes.




In order to test the validity of the linear approach, we also use the wake model to compute the bispectrum, comparing both the analytic expressions and the numerically evaluated functions in certain limits.
 As with the power spectrum, we use compensation factors to remove unphysical growing modes on superhorizon scales.

Having presented calculations of the bispectrum using both the (linear) perturbation based approach and the (non-linear) wake model approach, the general expression for the perturbation theory bispectrum is compared to the gravitational bispectrum in order to assess the possibility of using this signal to test for the presence of cosmic strings. By analogy to calculation for the matter bispectrum induced by self-ordering scalar fields \cite{Jaffe:1993tt,Figueroa:2010zx}, one may anticipate our conclusion that the cosmic string matter bispectrum is subdominant.


\subsection{UETC approach}

\subsubsection{General formalism}
We define the source bispectrum $\be$ by 
\ben
 \langle\Theta_+(\bk_1,\eta_1) \Theta_+(\bk_2,\eta_2)\Theta_+(\bk_3,\eta_3) \rangle
 =
 \phi_0^6 \be_+(k_1,k_2,k_3,\eta_1,\eta_2,\eta_3) (2\pi)^3 \delta^3(\bk_1+\bk_2+\bk_3).
\een
The source bispectrum is dimensionless in Fourier space, and so for a scaling source it can be expressed as a function of five dimensionless combinations of the six arguments $k_a$, $\eta_a$ ($a=1,2,3$). For example, one could take
$
x_a = k_a\eta_a, 
$
and chose any two of the combinations
\ben
r_1 = \frac{\eta_2}{\eta_3},  \quad r_2 = \frac{\eta_3}{\eta_1},  \quad r_3 = \frac{\eta_1}{\eta_2}.
\een
Note that $r_1r_2r_3 = 1$. 
It can also be convenient to define
$
\ka_{ab} = -\bk_a \cdot \bk_b,
$
and to use the dimensionless combinations $z_a$ defined by 
\ben
z_1^2 = \ka_{23}\eta_2\eta_3, \;
z_2^2 = \ka_{31}\eta_3\eta_1, \;
z_3^2 = \ka_{12}\eta_1\eta_2. 
\een
The fact that $z_a$ can be imaginary does not cause problems in practice.

The general form of the source bispectrum is not easy to guess, but we expect it to be strongly peaked  inside the horizon ($x_a \gg 1$) near  $r_a = 1$, by analogy with the source two-point function, and white noise outside the horizon.  Outside the horizon, we may use a similar argument to that described in Appendix \ref{sec:AppA} for the case of the power spectrum and infer that the three point correlations die off as the second power of the ratios of the earlier time to the later time.  Hence, a reasonable model is 
\ben
\label{e:SouBisFor}
\be_+(z_1,z_2,z_3,r_1,r_2) = 
\left\{
\ba{cc}
\displaystyle \frac{3E_+^{(3)}(0) }{\left(\frac{r_2}{r_1}\right)^{2} + \left(\frac{r_3}{r_2}\right)^{2} + \left(\frac{r_1}{r_3}\right)^{2} } , & (x_a \ll 1), \\[12pt]
E_+^{(3)}(z_1,z_2,z_3)\exp\left( -\sum_a z_a^2\ln(r_a)^2/2A^2 \right), & (x_a \gg 1) .
\ea
\right.
\een
We combine these regimes and without loss of generality assume that $\eta_1 < \eta_2,\eta_3$ to write
\ben
\label{e:SouBisFor2}
\be_+(z_1,z_2,z_3,r_1,r_2) = \frac{E_+^{(3)}(z_1,z_2,z_3)}{(r_3/r_2)^2}\exp\left( -\sum_a z_a^2\ln(r_a)^2/2A^2 \right)\,,
\een
i.e. we multiply the small scale source bispectrum by $1/(r_3/r_2)^2$ to obtain the correct large scale behaviour.
Substitution of the solution \eqref{deltaNsol} into \eqref{eq:bispdef} gives
\begin{align}
\label{eq:LinearMatterBisp}
B(k_1,k_2,k_3,\eta)
=\ep^3  \phi_0^6\int_{\eta_i}^{\eta} \int_{\eta_i}^{\eta} \int_{\eta_i}^{\eta}\Pi_{a=1}^3 [d\eta_a] \Pi_{a=1}^3
\left[ \GG_c(k_a;\eta,\eta_a)\gamma_c(k_a,\eta_a)\right]   \be_+(z_1,z_2,z_3,r_1,r_2) \,,\nonumber
\end{align}
where we have included the compensation factor $\ga_c$.

\subsubsection{Gaussian model for cosmic string \uetc3}
Following a similar calculation to that presented in \cite{HRS09x1,HRS09x2,ReganShellard09} and in Section~\ref{subsubsec:gaussian}, we find (with $\bk_3=-\bk_1-\bk_2$ imposed)
\begin{align*}
\be_+(k_1,k_2,k_3,\et_1,\eta_2,\eta_3) 
 \approx {8\bar\mu^3} {\Ll}\int d\sigma_{12}d\sigma_{13}\langle \dot{\bXs}_{1}^2 \dot{\bXs}_{2}^2 \dot{\bXs}_{3}^2\rangle\exp\left(-\frac{\sum_{a<b}\kappa_{ab}(\Gamma_{ab}(\eta_1)-\vba^2 \eta_{ab}^2)}{6}\right), 
\end{align*}
where we have adopted the shorthand notation $\sigma_{ab}=\sigma_a-\sigma_b$, $\bXs_{a}=\bXs(\sigma_a,\eta_a)$, $\Gamma_{ab}=\Gamma(\sigma_a-\sigma_b)$ and $\kappa_{ab}=-\bk_a \cdot \bk_b$. 
Suppressing the dependence on $\eta$, the individual correlators are given in the Gaussian approximation by 
\bea
\langle \dot{\bXs}_{1}^2 \dot{\bXs}_{2}^2 \dot{\bXs}_{3}^2\rangle
&=&
V^3(0)+2 V(0) V^2(\sigma_{12})+\frac{8}{9}V(\sigma_{12})V(\sigma_{13})V(\sigma_{23}), \nonumber\\
\exp\left(-\frac{1}{6}{\sum_{a<b}\kappa_{ab}\Gamma_{ab}^2(\eta')}\right)
&\approx&
\exp\left(-\frac{\overline{t}^2}{6}(k_2^2\sigma_{12}^2+\Ktwo^2\sigma_{13}^2)\right)\,, \nonumber\\
\exp\left(-\frac{\vba^2}{6}\sum_{a<b}\kappa_{ab}\eta_{ab}^2\right) 
&\approx&
\exp\left(-\frac{\vba^2}{6}(k_2^2 \eta_{12}^2+\Ktwo^2\eta_{13}^2)\right),
\eea
where $\Ktwo=\sqrt{k_2^2k_3^2-\kappa_{23}^2}/k_2$. In writing these expressions we assume that $k_2\gtrsim k_1,k_3$.
Approximating $V(\sigma)=\vba^2/2$ (with $V(0)=\vba^2$), we find that 
\ben
\be_+(k_1,k_2,k_3,\et_1,\eta_2,\eta_3) 
\approx
 \frac{\beta_0}{\xi^2} \frac{{1}}{\sqrt{k_2^2k_3^2-\kappa_{23}^2}}
\erf\left(\frac{\tba k_2 \xi}{\sqrt6}\right)\erf\left(\frac{\tba \Ktwo \xi}{\sqrt{6}}\right)
\exp\left(-\frac{\vba^2}{6}(k_2^2 \eta_{12}^2+\Ktwo^2\eta_{13}^2)\right)\,,\nonumber
\een
where we set $\be_0 = 8\bar\mu^3 ({29\pi}/{3}) ({\vba^6}/{\tba^2 })$. We assume $\eta_1< \eta_{2,3}$ and may, therefore, set $\xi={\rm{min}}(\xi(\eta_i))=\alpha \eta_1$ (see Section~\ref{subsubsec:gaussian}). 

Multiplying by $\eta_1^4/(\eta_2^2 \eta_3^2)$ we obtain the correct large scale behaviour (for $\eta_1<\eta_{2,3}$).
We note that the factor $k_2^2 k_3^2 -\kappa_{23}^2$ is proportional to the square of the area of the triangle formed by the three wavevectors $\cA$.  Indeed, 
\bea
4\cA^2 &=& k_2^2 k_3^2 -\kappa_{23}^2 
= \ka_{12}\ka_{23} + \ka_{23}\ka_{31} + \ka_{31}\ka_{12}
= \left(2(k_1^2k_2^2 + k_2^2k_3^2 + k_3^2k_1^2) - k_1^4  - k_2^4  - k_3^4\right)/4\,. \nonumber 
\eea

\subsection{Wake model approach}
The form of the matter bispectrum using the non-linear wake model approach is easily inferred from equations \eqref{eq:delta1}, \eqref{eq:bispdef}, and the discussion in Section~\ref{sec:wakemodelpower},
\bea\label{eq:wakemodelbisp}
B(k_1,k_2,k_3,\eta)=&&\int_{\eta_i}^{\eta}d\eta_i\frac{\nu}{\eta_i^4}(2\pi\xi(\eta_i)^2\bar{\sigma}_w(\eta,\eta_i))^3 \Pi_{a=1}^3 \gamma_c(k_a,\eta_i) \int \frac{d\hat{\bn}}{4\pi}\Pi_{a=1}^3\left[\frac{ J_1(k_{a\perp}\xi(\eta_i)}{k_{a\perp}\xi(\eta_i)}\right] \,.\nonumber\\
\eea
Using equations \eqref{e:nuDef} and the expressions in Section \ref{subsec:wakemodel} we may rewrite this in the form
\bea\label{eq:wakemodelbisp2}
B(k_1,k_2,k_3,\eta)=&&(8\pi G\mu)^3 \frac{\vba^4}{\tba^2 \alpha}\frac{64 \pi}{125}\eta^6\int_{\eta_i}^{\eta}\frac{d\eta_i}{\eta_i^4}\left[\Pi_{a=1}^3 \gamma_c(k_a,\eta_i)  \right]  \int \frac{d\hat{\bn}}{4\pi}\Pi_{a=1}^3\left[\frac{ J_1(k_{a\perp}\xi(\eta_i)}{k_{a\perp}}\right]\,.\nonumber\\
\eea

\subsection{Analytic limits}

\subsubsection{UETC analytic limits}
The bispectrum of string-induced matter perturbations may be written
\bea\label{eq:bispUETC}
B(k_1,k_2,k_3,\eta)\approx&&(8\pi G\mu)^3 \frac{29\pi}{375}\frac{\vba^6}{\alpha^2 \tba^2}\int_{\eta_i}^{\eta} \frac{d\eta_1}{\eta_1^3} \int_1^{\eta/\eta_{i}} \frac{dr_2}{r_2^3} \int_1^{\eta/\eta_{i}} \frac{d\tilde{r}_3 }{\tilde{r}_3^3 }\erf\left(\frac{\tba k_2 \alpha \eta_1}{\sqrt6}\right)\erf\left(\frac{\tba \Ktwo \alpha \eta_1}{\sqrt{6}}\right)\nonumber\\
&&\times \exp\left(-\frac{\vba^2}{6}(k_2^2 \eta_{1}^2 (\ln \tilde{r}_3)^2+\Ktwo^2\eta_{1}^2 (\ln r_2)^2)\right)\gamma_c(k_1,\eta_1)\gamma_c(k_2,\eta_1 \tilde{r}_3)\gamma_c(k_3,\eta_1 r_2)\nonumber\\
&&\times \tilde{\GG}_c(k_1;\eta,\eta_1)\tilde{\GG}_c(k_2;\eta,\eta_1 \tilde{r}_3)\tilde{\GG}_c(k_3;\eta,\eta_1 r_2) \frac{\eta^6}{k_2 \Ktwo} \,,
\eea
where $\tilde{r}_3=1/r_3$, and we set $\tilde{\GG}_c(k;\eta,\eta_i)=\GG_c (k;\eta,\eta_i)5\eta_i/\eta^2$, a function which tends to one in the deep matter era. 
In the following limits we shall assume that $k_3 \lesssim k_1 = k_2=k$, and define 
\ben
\Ktwo=k_3\sqrt{1-k_3^2/(4 k^2)}.
\een 
We assume the range of integration ranges from $\eta_1\in [\eta_{\rm eq},\eta]$, where $\eta\gg \eta_{\rm eq}$, such that 
we may assume $\tilde{\GG}_c=1$. We shall consider firstly the small scale limits, and then consider the squeezed limit result, as well as, the result for the large scale equilateral limit.

The analytic estimates may be computed in a similar fashion to those of the power spectrum, as detailed in Appendix~\ref{sec:AppC}. 
\para{Small scales, $k_2,\Ktwo \gg \eta_{\rm eq}^{-1}$}
In this regime we may neglect the compensation factors and set the error functions to unity. This results in the small scale bispectrum result,
\bea\label{eq:smallscale}
\BpostEq(k_1,k_2,k_3,\eta)\approx  (8\pi G\mu)^3 \frac{29\pi^2}{1000}\frac{\vba^4}{\alpha^2 \tba^2} \frac{\eta^6}{(k_2 \Ktwo)^2 \eta_{\rm eq}^4}\,.
\eea
In particular, in the equilateral limit (for which $\Ktwo=\sqrt{3}k/2$), we observe that $\BpostEq^{\rm equil}\propto \eta^6/k^4$ in the small scale limit. For the (slightly) squeezed limit where $k_2\gg k_3$, but $k_3\gg \eta_{\rm eq}^{-1}$, this result implies that $\BpostEq^{\rm squ}\propto \eta^6/(k k_3)^2$. However, one usually pictures the squeezed limit for which $\eta^{-1} \ll k_3\ll \eta_{\rm eq}^{-1} \ll k_2$, which as we shall see next has different qualitative behaviour.

\para{Special case: squeezed limit, $k\gg \eta_{\rm eq}^{-1}$, $ \eta^{-1} \ll \Ktwo \ll \eta_{\rm eq}^{-1} $}
In the time interval $\eta'\in [\eta_{\rm eq},\sqrt{18}/k_3]$ the compensation factor for the scale $k_3$ may not be neglected, as it is outside the horizon. 
With the compensation included, the bispectrum \eqref{eq:bispUETC} is approximated by
\bea
\BpostEq(k,k,k_3,\eta)\approx&& (8\pi G\mu)^3  \frac{29\pi}{6750} \frac{\vba^5}{\alpha \tba} \frac{k_3^2}{k^2} \eta^6 
\ln\left(\frac{\sqrt{18}}{k_3\eta_{\rm eq}}\right)^2 \propto \frac{k_3^2}{k^2}\,.
\eea 


\para{Special case: large scale equilateral limit, $ \eta^{-1}\ll k \ll  \eta_{\rm eq}^{-1}$}
To obtain the large scale equilateral result, we consider the integration limits $\eta_1\in [\sqrt{18} k^{-1},\eta]$ for which we may neglect the compensation factors. The regime $\eta_1\in [\eta_{\rm eq},\sqrt{18} k^{-1}]$ gives the same qualitative behaviour but is of smaller amplitude. We find
\bea
\BpostEq(k,k,k,\eta)\approx&& (8\pi G\mu)^3  \frac{29\pi^2}{750}\frac{\vba^4}{\alpha^2 \tba^2}\eta^6 \propto k^0\,.
\eea 
Therefore, in this regime we obtain the result that the matter bispectrum in the equilateral configuration does not scale with wavenumber at each redshift.

\subsubsection{Wake model analytic limits}
In order to validate the qualitative behaviour of the (linear) perturbation theory matter bispectrum, we consider the same analytic limits using the non-linear wake model -  employing equation~\eqref{eq:wakemodelbisp2} in our study.\footnote{We omit the general small scale limit for which an analytic result is not possible and include the specific case of an equilateral small scale limit.}


\para{Small scale equilateral limit, $k_i = k \gg \eta_{\rm eq}^{-1}$}
We parametrise the equilateral shape such that $\bk_1 =k(\sin(\pi/3),\cos(\pi/3),0)$, $\bk_2 =k(-\sin(\pi/3),\cos(\pi/3),0)$, with $\bk_3=-\bk_1-\bk_2$, such that 
\bea
k_{1\perp}^2/k^2 &&= 1-\sin^2\theta \sin^2 (\phi+\pi/3)\,, \quad
k_{2\perp}^2/k^2 = 1-\sin^2\theta\sin^2 (\phi-\pi/3)\,, \nonumber\\
k_{3\perp}^2/k^2 &&= 1-\sin^2\theta\sin^2 (\phi)\,,
\eea
where $d \hat{\bn}=\sin \theta d\theta d\phi$. Setting $\mu=\cos \theta$, the integrand in equation~\eqref{eq:wakemodelbisp2} peaks near the points on the $(\mu,\phi)$ plane given by $(1,\pi/2), (1,\pi/2\pm \pi/3)$. Clearly by symmetry considerations each of the peaks contributes equally. Therfore, expanding in the $\phi$ parameter near the peak at $\pi/2$, in the form $\phi=\pi/2+\epsilon$, we may write $k_{3\perp}/k=\sqrt{\epsilon^2 + \mu^2}$, $k_{1 \perp}/k\approx k_{2 \perp}/k \approx \sqrt{(3+\mu^2)/4}$. We may then obtain the result,
\begin{align}
&\int \frac{d\hat{\bn}}{4\pi} \Pi_{i=1}^3\left[\frac{J_1(k_{i\perp} \alpha \eta_i) }{k_{1\perp} \alpha \eta_i }\right] \approx 3\int \frac{d\mu d\epsilon}{4\pi} \frac{J_1(z\sqrt{\epsilon^2 +\mu^2})J_1(z\sqrt{3 +\mu^2}/2)^2     }{ z\sqrt{\epsilon^2 +\mu^2} (z\sqrt{3 +\mu^2}/2)^2   }  \nonumber\\
&\approx \frac{3}{4\pi} \left(\frac{J_1(\beta z)}{\beta z}\right)^2 \int d\mu d\epsilon \frac{  J_1 (z \sqrt{\epsilon^2 +\mu^2})  }{  z\sqrt{\epsilon^2 +\mu^2 }     }
\approx  \frac{3}{2}  \left(\frac{J_1(\beta z)}{\beta z}\right)^2 \int d\rho \frac{J_1(z\rho)}{z}\approx  \frac{3}{2}  \frac{(J_1(\beta z))^2}{\beta^2 z^4}\,,
\end{align}
where we set $z\equiv k\alpha \eta_{i}$, and where $\beta$ represents a constant of value between $3/4$ and $1$. On the final line of this expression we use the fact that $\int_0^z dx J_1(x)=1-J_0(z)\approx 1$ (for $z\gg 1$). Finally our analytic expression for the equilateral bispectrum takes the form
\begin{align}
\BpostEq(k,k,k,\eta)\approx  (8\pi G\mu)^3 \frac{\vba^4}{\tba^2 \alpha^2}\frac{24 \pi}{125 \beta^2} \frac{\eta^6}{\eta_{\rm eq}^4 k^4}\,.
\end{align}
The agreement with equation~\eqref{eq:smallscale} in the equilateral limit is immediately apparent, suggesting again that the linear perturbation theory result may be trusted for calculations of the three-point function.

\para{Special case: squeezed limit, $k\gg \eta_{\rm eq}^{-1}$, $ \eta^{-1} \ll \Ktwo \ll \eta_{\rm eq}^{-1} $}
In this regime we may again replace the $k_3$ dependent Bessel function using the small $x$ limit $J_1(x)\approx x/2$, to obtain the result
\ben
\int \frac{d\hat{\bn}}{4\pi} \Pi_{i=1}^3\left[\frac{J_1(k_{i\perp} \alpha \eta_i) }{k_{1\perp} \alpha \eta_i }\right] \approx \int \frac{d\hat{\bn}}{4\pi} \frac{J_1(k_{\perp} \alpha \eta_i)^2 }{(k_{\perp} \alpha \eta_i)^2 }=\int_{-1}^1 \frac{d\mu}{2}\frac{J_1(k\sqrt{1-\mu^2}\alpha \eta_i)^2 }{(k\sqrt{1-\mu^2} \alpha \eta_i)^2 }\approx \frac{1}{\alpha^2 \eta_i^2 k^2}\,.
\een 
Therefore, we find that
\bea
\BpostEq(k,k,k_3,\eta)=(8\pi G\mu)^3 \frac{\vba^4}{\tba^2}\frac{32 \pi}{1125}\frac{k_3^2}{k^2} \eta^6 \ln\left(\frac{\sqrt{18}}{k_3 \eta_{\rm eq}} \right)\,, 
\eea
which again agrees well with the perturbation theory result in this limit.

\para{Special case: large scale equilateral limit, $ \eta^{-1}\ll k \ll  \eta_{\rm eq}^{-1}$}
The calculation follows similar steps to the small scale equilateral limit, with the result in this case given by
\bea
\BpostEq(k,k,k,\eta)\approx  (8\pi G\mu)^3 \frac{\vba^4}{\tba^2 \alpha^2}\frac{24 \pi}{125 \beta^2} \eta^6 \propto k^0\,.
\eea
We conclude that the non-linear and linear models give the same matter bispectrum in all limits considered. As such we may use the bispectrum calculated using linear perturbation theory valid for all redshifts (and not just the deep matter era as studied analytically in this section) to investigate numerically whether this signal may be detectable or otherwise.

\subsection{Numerical evaluation and comparison to gravitational bispectrum}
\begin{figure}
\centering
\hspace{0.1cm}
{\includegraphics[width=0.44\linewidth]{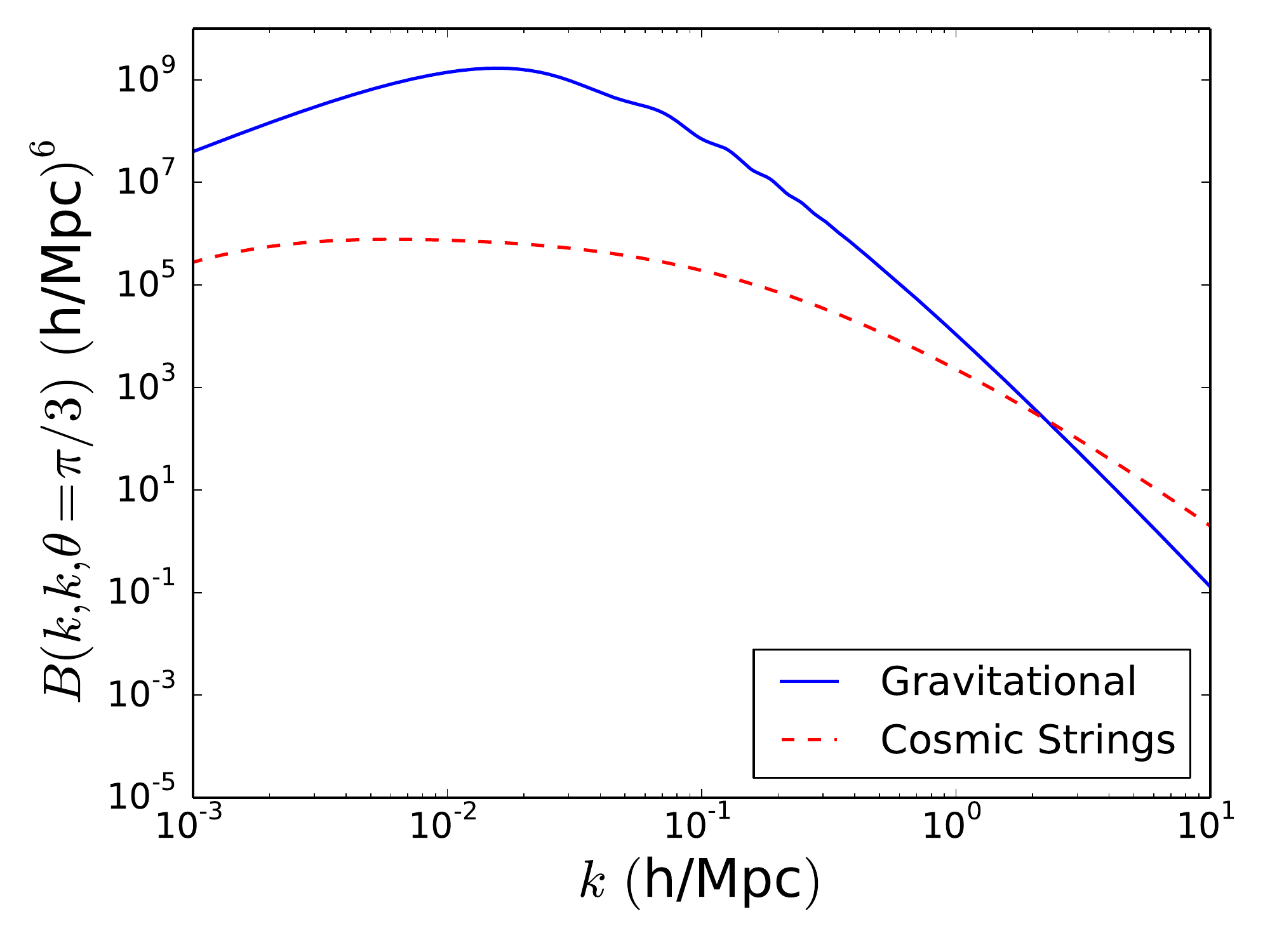}}
\hspace{0cm}
{\includegraphics[width=0.44\linewidth]{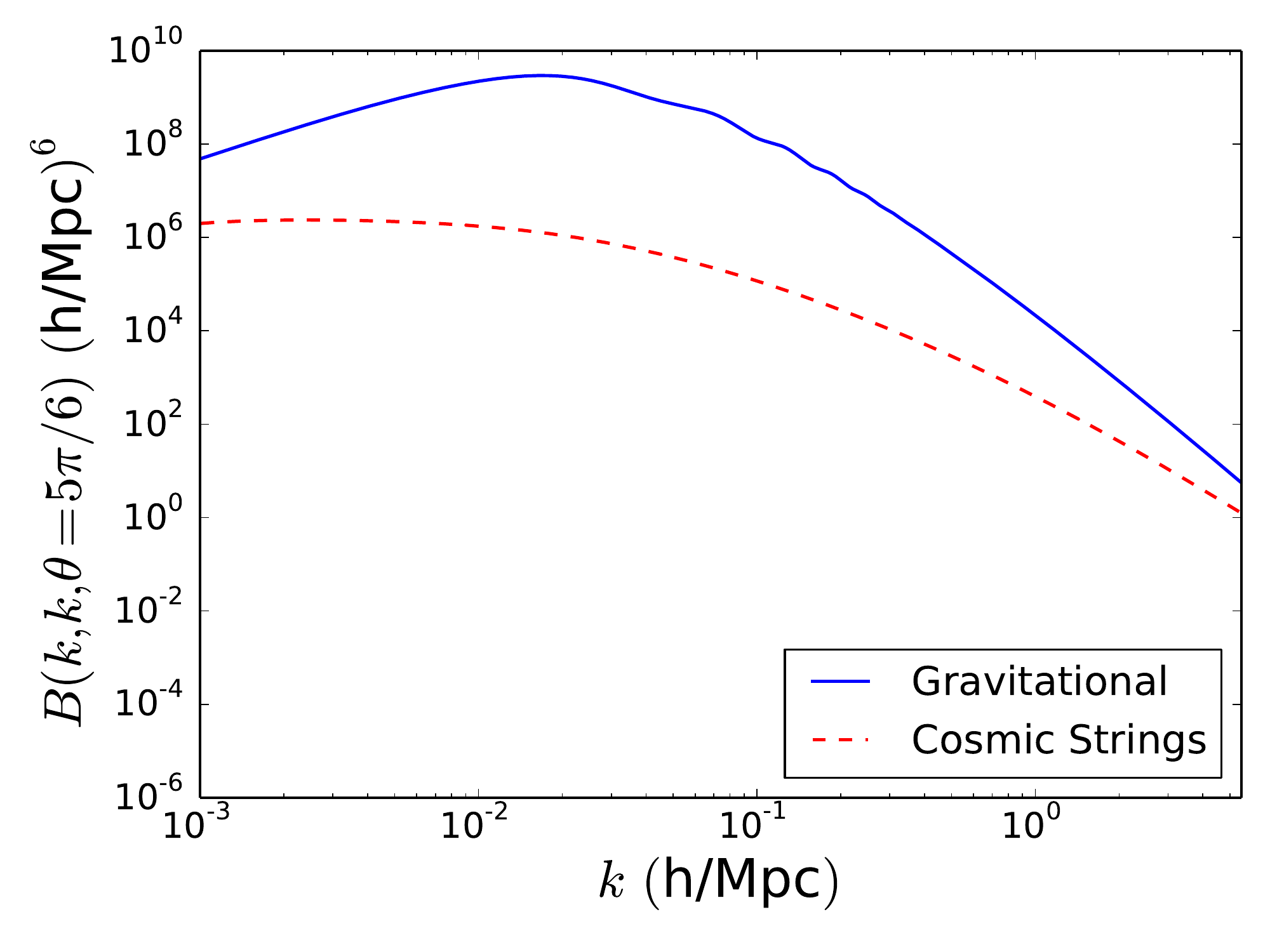}}
{\includegraphics[width=0.44\linewidth]{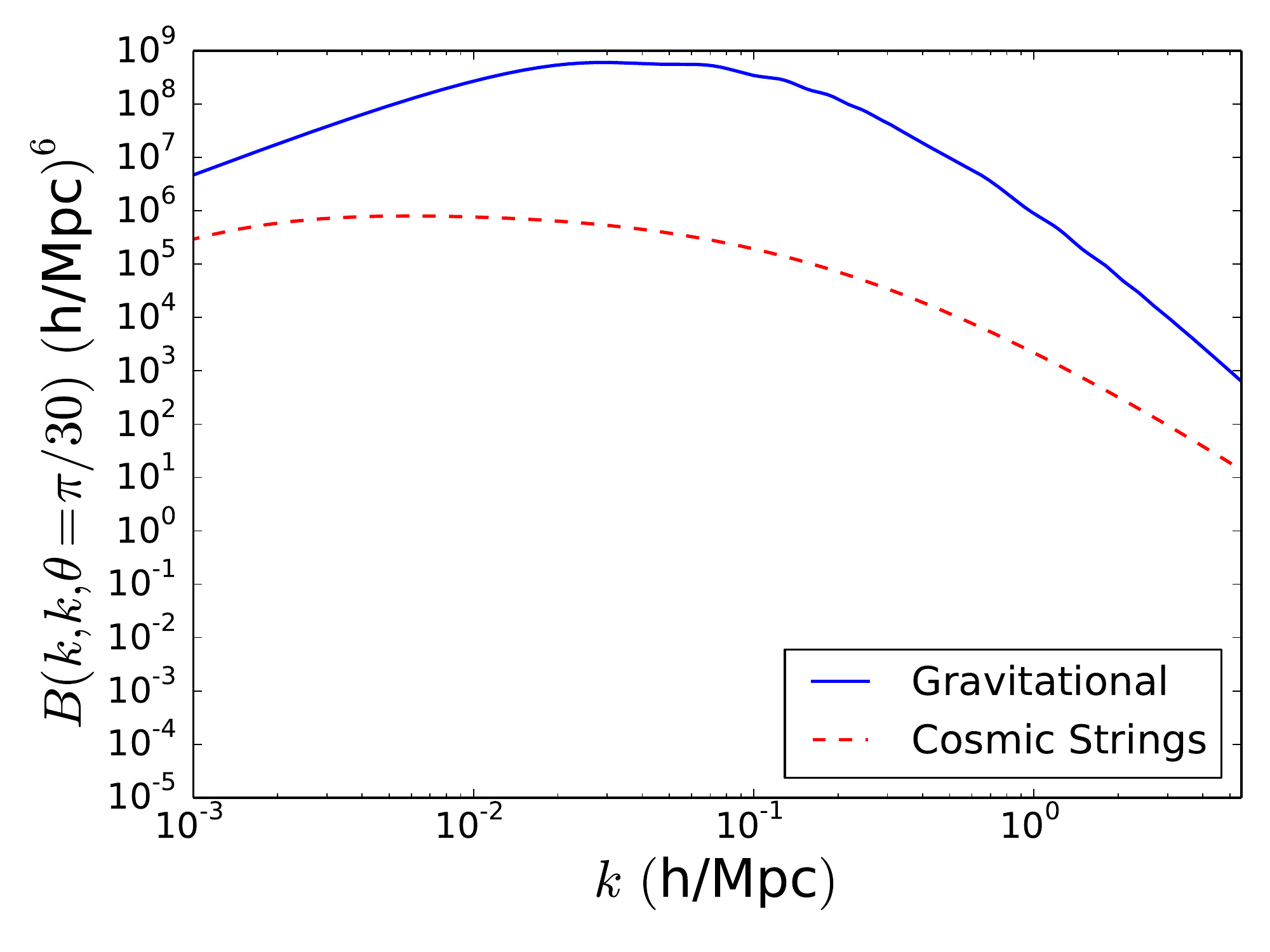}} 
\caption{Comparison of the gravitational bispectrum~\eqref{eq:bispgrav} with the matter bispectrum of cosmic strings for $k_1=k=k_2$ and $k_3=k\sin \psi$. Clockwise: first panel describes the comparison for $\psi=\pi/3$, i.e. equilateral configuration, second panel is for $\psi=5\pi/6$ (describing a folded configuration), third panel is for $\psi=\pi/30$ (squeezed configuration). The plots are computed for redshift $z=0$ and for string tension parameter $G\mu=1.1\times 10^{-6}$. Note that string bispectra scale as $(G\mu)^3$.}
\label{fig:various2}
\end{figure}

We obtain numerical values for the bispectrum~\eqref{eq:bispUETC} using the numerical solutions for the Green's function and the VOS model, 
in just the same way as the power spectrum. We consider the bispectrum for $k_1=k_2=k$ and $k_3<k$. We compare to the gravitational bispectrum given (at tree level) by
\bea\label{eq:bispgrav}
B^{\rm grav}(k_1,k_2,k_3)=2 F_2(\bk_1,\bk_2)\tilde{P}(k_1)\tilde{P}(k_2) + 2\,\,\rm perms\,,
\eea
where the kernel $F_2$ is given by \cite{Bernardeau:2001qr}
\bea
F_2(\bk_1,\bk_2)=\frac{5}{7}+\frac{1}{2}\frac{\bk_1\cdot\bk_2}{k_1 k_2}\left(\frac{k_1}{k_2}+\frac{k_2}{k_1}\right)+\frac{2}{7}\left(\frac{\bk_1\cdot\bk_2}{k_1 k_2}\right)^2\,.
\eea
Setting $k_3=k\sin \psi$, in Figure~\ref{fig:various2} we plot the comparison between the gravitational bispectrum and the cosmic string matter bispectrum for $\psi=\pi/3$ (equilateral limit), $\psi=5\pi/6$ (to probe the folded limit) and $\psi=\pi/30$ (to probe the squeezed limit). We find that the gravitational bispectrum for $k\lesssim 3 h {\rm Mpc}^{-1}$ dominates the cosmic string bispectrum. These plots are produced for $G\mu = 1.1 \times 10^{-6}$, and noting that the cosmic string bispectrum is proportional to $(G\mu)^3$ we infer that, at current observational limits (about $3\times 10^{-7}$ in the field theory scenario and $10^{-7}$ in the Nambu-Goto scenario), the cosmic string bispectrum lies several orders of magnitude below the gravitational bispectrum. Considering the redshift behaviour, one deduces that the gravitational bispectrum scales as $1/(1+z)^4$, while the cosmic string bispectrum scales as $1/(1+z)^3$. Therefore, while the cosmic string bispectrum increases (relatively) by an order of magnitude at $z\sim 10$, it still lies far below the gravitational bispectrum, and so is unlikely to be detectable. 

\section{Discussion and conclusions}\label{sec:conclusions}
In this paper we have calculated the bispectrum of cosmic string induced matter fluctuations using two different approaches. 
The first uses linear perturbation theory and integrates a three-point unequal time correlator (UETC3) of the energy momentum tensor with the Green's function for the matter perturbations. The UETC3 uses a Gaussian model for the string position and velocity correlators, which are compared with those in the widely-used unconnected segment model (USM). We show that they have a similar form, and result in very similar power spectra. 
The second approach uses the Zel'dovich approximation to compute the density fluctuation induced on a planar surface behind a moving straight string. This wake model  captures some of the non-linearity and non-Gaussianity of the perturbations induced by strings.

We find that, provided compensation factors are used to maintain energy conservation on large scales, the two models give the same shape and growth rate for the matter bispectrum. 
The equilateral limit decreases as $k^{-4}$ on scales less than the horizon at matter-radiation equality, while in the squeezed limit $k_3\ll k_1=k_2 = k$, with $k_3$ representing a large scale mode at matter-radiation equality, the bispectrum scales as $k_3^2/k^2$, vanishing in the very squeezed limit. As the triangle becomes less squeezed and $k_3$ approaches smaller scales, the bispectrum reverts to a $1/(k_3^2 k^2)\sim 1/k^4$ scaling as in the equilateral case. The large scale behaviour is approximately scale invariant, agreeing qualitatively with results for self-ordering scalar fields in this limit. 

In all cases the bispectrum is proportional to $(G\mu)^3$ and grows as the scale factor cubed. By contrast the gravitational bispectrum grows as the square of the scale factor. However, despite the relative increase of magnitude of the cosmic string signal with redshift, the signal still lies at least an order of magnitude below the gravitational bispectrum, at observable redshifts, and observable scales. A more quantitative comparison to the gravitational bispectrum reveals that the cosmic string signal is expected to lie several orders of magnitude below at current observational limits, at redshifts $z\lesssim 100$, and wave numbers $k\lesssim 1 h/{\rm Mpc}$. The results indicate that future searches for cosmic strings using higher order correlators of large scale structure are unlikely to prove competitive with other probes such as the CMB.

\section*{Acknowledgements}
We would like to thank Anastasios Avgoustidis, Jon Urrestilla and Martin Kunz for helpful comments and suggestions on a draft of this work.
DR acknowledges support from the Science and Technology Facilities Council
(grant number ST/I000976/1). DR is also supported by funding from the European Research Council under the European Union's Seventh Framework Programme (FP/2007-2013) / ERC Grant Agreement No. [308082]. 
This work was undertaken on the COSMOS Shared Memory system at DAMTP, University of Cambridge operated on behalf of the STFC DiRAC HPC Facility. This equipment is funded by BIS National E-infrastructure capital grant ST/J005673/1 and STFC grants ST/H008586/1, ST/K00333X/1.

\label{acknow}

\appendix

\section{Superhorizon scale behaviour}\label{sec:AppA}
At any scale $R$, the volume in each horizon volume can be regarded as a random process
$\rho_R(\bx,\eta)=\bar{\rho}_R(\eta)+\delta\rho_R(\bx,\eta)$.
We consider a relative fluctuation of $\mathcal{O}(1)$ on each horizon scale such that $\sqrt{\langle\delta\rho_R(\bx,\eta)^2\rangle}=\left(\frac{\eta}{R}\right)^{3/2}\bar{\rho}_R(\eta)$. Therefore, the two point correlator for $\eta_1<\eta_2$ on such scales obeys
\begin{align}
\langle \delta \rho_R(\bx,\eta_1)\delta \rho_R(\bx,\eta_2)\rangle = \left(\frac{\eta_1}{R}\right)^{3/2}\bar{\rho}_R(\eta_1) \times \left(\frac{\eta_2}{R}\right)^{3/2}\bar{\rho}_R(\eta_2)\times \left(\frac{\eta_1}{\eta_2}\right)^{3/2}=\frac{\eta_1^3}{R^3} \overline{\rho}_R(\eta_1)\overline{\rho}_R(\eta_2)\,.
\end{align}
Expressing this equation in Fourier space we find that for white noise on super-horizon scales,
\begin{align}
\langle \delta \rho_R(\bx,\eta_1)\delta \rho_R(\bx,\eta_2)\rangle&=\int_0^{2\pi/R} d^3 k \langle \delta \rho_R(\bk,\eta_1)\delta \rho_R(\bk,\eta_2)\rangle\propto \langle \delta \rho_R(\bk,\eta_1)\delta \rho_R(\bk,\eta_2)\rangle/R^3 \nonumber\\
\implies  \langle \delta \rho_R(\bk,\eta_1)\delta \rho_R(\bk,\eta_2)\rangle&\propto \eta_1^3 \bar{\rho}(\eta_1) \bar{\rho}(\eta_2)\,.
\end{align}
Since $\bar{\rho}(\eta_i)\propto 1/\eta_i^2$ (as must be satisfied for a scaling network of cosmic strings) we infer that the super-horizon behaviour of the unequal time correlator must satisfy
\begin{align}\label{eq:densitysuperhor}
\langle \delta \rho_R(\bk,\eta_1)\delta \rho_R(\bk,\eta_2)\rangle \propto \frac{\eta_1^3}{\eta_1^2 \eta_2^2}= \frac{1}{\sqrt{\eta_1 \eta_2}}\left(\frac{\eta_1}{\eta_2}\right)^{3/2}\,.
\end{align}
While this description is strictly only valid for the density component of the stress energy tensor, i.e. $\Theta_{00}$, we note from energy momentum conservation that on super horizon scales
\begin{align}
\dot{\Theta_{00}}+\frac{\dot{a}}{a}\Theta_+ \rightarrow 0\,,
\end{align}
which implies that on super horizon scales super horizon form of $\langle \Theta_+ \Theta_+^*\rangle\propto $ is also given by the right hand side of \eqref{eq:densitysuperhor}.
Using the notation of Section~\ref{sec:powerspecSEC} this implies that $C_+(k,\eta_1,\eta_2)\propto \left({\eta_1}/{\eta_2}\right)^{3/2}$ for $\eta_1<\eta_2$.


\section{Power Spectrum Analytic Limits}\label{sec:AppC}
\subsection{UETC analytic limits}
In order to gain an understanding for the qualitative behaviour of the matter power spectrum, we present analytic expressions in the limit of a matter dominated universe. We restrict the range for the integrals in equation~\eqref{e:SubPSdef} to $[\eta_{\rm eq},\eta]$ (where $\eta\gg \eta_{\rm eq}$). We may then approximate the Green's function by
\ben
\GG_c(k;\eta,\eta')\approx\frac{\eta^2}{5\eta'}\,.
\een
In this regime the compensation factor may be written as $\gamma_c(k,\eta_1)=1/(1+18/(k^2 \eta_1^2))$.
Changing variables to $(z,r)$ gives
\ben
\PpostEq(k,\eta)\approx \ep^2 \phi_0^4\frac{k\eta^4}{25} \int_{\zEq}^z dz' z' \int_1^\infty \frac{dr}{r} \frac{C_+(z',r)}{{z'}^3}\frac{1}{1+18r/z'^2} \frac{1}{1+18/(z'^2 r)} 
\een

\para{Superhorizon scales, $k \lesssim \eta^{-1}$}
For superhorizon scales, the UETC is $z$ independent, and we may write
\bea
\PpostEq(k,\eta) &\approx& \ep^2 \phi_0^4\frac{k\eta^4}{25} \int_{\zEq}^z {dz'}\frac{{z'}^2}{18^2} \int \frac{dr}{r} \frac{2\bar{E}_+^\text{m}}{r^\frac{3}{2} + r^{-\frac{3}{2}}}\approx \ep^2 \phi_0^4\frac{2\pi \bar{E}_+^\text{m}}{225} \frac{1}{18^2} {k^4 \eta^7}
\eea
From equations~\eqref{e:GenForUETC} and~\eqref{e:EplusAvals} we infer that $\bar{E}_+^m \approx 28 \bar{\mu}^2 \vba^4/(3 \alpha)$, where we use the limit for the error function
$
{\rm erf}(x)\approx {2x/\sqrt{\pi}}\,, $ for  $x\ll 1\,.
$
Hence, we obtain the result for the superhorizon behaviour
\bea
\PpostEq(k,\eta) &\approx& (8\pi G\mu)^2 \frac{14 \pi}{675} \frac{1}{18^2}\frac{\vba^4}{\alpha} k^4 \eta^7 \propto k^4\,.
\eea

\para{Large scales, $ \eta^{-1} \ll k \ll \etaEq^{-1}$}
For perturbations on length scales smaller than the horizon, but still large compared with the horizon at matter-radiation equality, we restrict the range of the $z$ integration to $z'\in [\sqrt{18},z]$, such that the compensation factor terms may be neglected (the contribution from the regime $z'\in [\zEq,\sqrt{18}]$ has the same qualitative behaviour but is of smaller amplitude). In this regime we may also approximate the error function by unity. We then obtain the large scale power spectrum,
\bea
\PpostEq(k,\eta) &\approx& 
\ep^2  \phi_0^4\frac{k\eta^4}{25}
 \int_{\sqrt{18}}^z \frac{dz'}{{z'}^2} \int \frac{dr}{r}  \frac{E_+^\text{m}(z')}{z'}e^{-\frac{{z'}^2\ln(r)^2}{2A^2}} \approx (8\pi G\mu)^2 \frac{7\pi}{150}\frac{\vba^3}{\tba \alpha^2} \frac{k\eta^4}{18^{3/2}}\propto k
\eea
Hence we see that inside the horizon, the growing mode of the perturbation induced by the string takes the same scale-invariant form as inflation-induced perturbations, proportional to $k$.

\para{Small scales, $\etaEq^{-1} \ll k $}
For small scales, perturbations after matter-radiation equality are sourced by strings already inside the horizon. In this regime the compensation factors are unimportant and the power spectrum takes the form
\bea
\PpostEq(k,\eta) &\approx& 
\ep^2 \phi_0^4 \frac{k\eta^4}{25}
 \int_{\zEq}^z \frac{dz'}{{z'}^2} \int \frac{dr}{r}  \frac{E_+^\text{m}(z')}{z'}e^{-\frac{{z'}^2\ln(r)^2}{2A^2}}  
= (8\pi G\mu)^2 \frac{7\pi}{150}\frac{\vba^3}{\tba \alpha^2} \frac{\eta^4}{k^2 \eta_{\rm eq}^3}\propto k^{-2}.
\eea
For large $k\etaEq$ we see the characteristic $k^{-2}$ behaviour characteristic of strings, and consistent with their perturbations being in the form of approximately planar wakes \cite{Albrecht:1991br}.  We will show in more detail later in this section how string wakes produce a $k^{-2}$ spectrum.

\subsection{Wake model analytic limits}
We use equation~\eqref{eq:powerwake} to calculate the power spectrum using the wake model approach in this section. We again limit the time integration from $\eta_{\rm eq}$ to $\eta$. We expect the analytic estimates for both approaches to be both qualitatively and quantitatively similar (though given the approximations used we require a numerical evaluation of the Green's functions in the perturbation theory approach to get accurate quantitative results).

\para{Superhorizon scales, $k \lesssim \eta^{-1}$}
In the regime $x\ll1$ we may make the replacement $J_1(x)\approx x/2$. On super horizon scales we may approximate the compensation factor as $\gamma_c(k,\eta_1)\approx k^2 \eta_1^2/18$. We may then write the power spectrum in the form
\ben
\PpostEq(k,\eta) \approx \frac{(2\pi)^2}{18^2} \nu \alpha^4 k^4\int_{\eta_{\rm eq}}^\eta d\eta_i \bar{\sigma}_w^2(\eta,\eta_i)\approx \frac{(8\pi G\mu)^2}{18^2} \frac{8}{75}\frac{\vba^3}{\tba} k^4 \eta^7 \propto k^4\,.
\een
Thus we see that this has the same qualitative behaviour as for the perturbation theory result. Comparing the coefficients we find that they agree up to factors of order unity.

\para{Large scales, $ \eta^{-1} \ll k \ll \etaEq^{-1}$}
We split the integral into the regime where the compensation factor is important ($\eta_i\in [\eta_{\rm eq},\sqrt{18}k^{-1}]$) from where it may be neglected ($\eta_i\in [\sqrt{18}k^{-1},\eta]$). In the former regime we again use the replacement $J_1(x)\approx x/2$, while in the latter we exploit the property that $\int_0^{\infty}dx J_1(x)^2/x =1/2 $ by rewriting
\begin{align}\label{eq:intJ1}
\int_{-1}^1 d\mu \frac{(J_1(k \alpha \eta_i\sqrt{1-\mu^2}))^2}{k^2(1-\mu^2)}&=2\int_{0}^{k\alpha \eta_i}dx \frac{J_1(x)^2}{x}\frac{(\alpha \eta_i)^2}{(k \alpha \eta_i)^2\sqrt{1-(x/(k\alpha \eta_i)^2)}}\nonumber\\
&\approx 2\int_{0}^{\infty}dx \frac{J_1(x)^2}{x}\frac{1}{k^2}=\frac{1}{k^2}\,.
\end{align}
where in the second line we assume that $k\alpha \eta_i \gg 1$. Therefore, we find that
\bea
\PpostEq(k,\eta) &\approx& (8\pi G\mu)^2 \frac{16}{25} \frac{\vba^3}{\tba} \eta^4 \left[ \int_{\eta_{\rm eq}}^{\sqrt{18}k^{-1}}d\eta_i  \frac{k^4}{18^2}\frac{\eta_i^2}{4}+\int^{\eta}_{\sqrt{18}k^{-1}} \frac{d\eta_i}{\eta_i^4}\frac{1}{2 k^2 \alpha^2} \right] \nonumber\\
&\approx&  (8\pi G\mu)^2 \frac{4}{75\sqrt{18}} \frac{\vba^3}{\tba}\left(1+\frac{1}{9\alpha^2}\right)  \eta^4 k \propto k\,.
\eea

\para{Small scales, $\etaEq^{-1} \ll k $}
In this regime we may neglect the compensation factor. We make use of equation~\eqref{eq:intJ1} to approximate the power spectrum by
\bea
\PpostEq(k,\eta) &\approx& (8\pi G\mu)^2 \frac{8}{75} \frac{\vba^3}{\tba \alpha^2} \frac{\eta^4}{k^2\eta_{\rm eq}^3} \propto 1/k^2  \,.
\eea
We again observe the $1/k^2$ behaviour of the power spectrum, which is a characteristic signature of the wake-like structure induced by cosmic strings.

In summary, the non-linear wake model and the linear perturbation theory based approach produce the same behaviour in the super horizon, large scale and small scale regimes. In addition, it is clear that the compensation term is an important large scale effect which needs to be incorporated in both approaches.

%

\bibliography{bibli,CosmicStrings}
\end{document}